\documentclass[5p,times]{elsarticle}

\usepackage{amsmath}
\usepackage{algorithm}
\usepackage{algpseudocode}
\usepackage{multirow}
\usepackage{comment}
\usepackage{subcaption}
\usepackage{amsfonts}
\usepackage{hyperref}
\usepackage{amsthm}

\usepackage{hyperref}

\usepackage{xcolor}
\title{\texttt{parGeMSLR}: A Parallel Multilevel Schur Complement Low-Rank Preconditioning and Solution Package for General Sparse Matrices}

\author[1]{Tianshi Xu\corref{cor1}}
\ead{xuxx1180@umn.edu}
\author[2]{Vassilis Kalantzis}
\ead{vkal@ibm.com}
\author[3]{Ruipeng Li}
\ead{li50@llnl.gov}
\author[4]{Yuanzhe Xi}
\ead{yxi26@emory.edu}
\author[5]{Geoffrey Dillon}
\ead{DILLONG@math.sc.edu}
\author[1]{Yousef Saad}
\ead{saad@cs.umn.edu}

\address[1]{Department of Computer Science and Engineering, University of Minnesota, Minneapolis, MN 55455}
\address[2]{Thomas J. Watson Research Center, IBM Research, Yorktown Heights, NY 10598}
\address[3]{Center for Applied Scientific Computing, Lawrence Livermore National Laboratory, P.O. Box 808, L-561, Livermore, CA  94551}
\address[4]{Department of Mathematics, Emory University, Atlanta, GA 30322}
\address[5]{Department of Mathematics, University of South Carolina, Columbia, SC 29208}
\cortext[cor1]{
The work of the first and the last author was supported by the National Science Foundation (NSF) grant DMS-1912048. The work of the fourth author was supported by the NSF grant OAC-2003720. This work was performed under the auspices of the U.S. Department of Energy by Lawrence Livermore National Laboratory under Contract DE-AC52-07NA27344 (LLNL-JRNL-830724)
}

\begin{document}

\begin{abstract}
This paper discusses \texttt{parGeMSLR}, a C++/MPI software library for the solution of sparse 
systems of linear algebraic equations via preconditioned Krylov subspace methods in distributed-memory computing environments. The preconditioner implemented in \texttt{parGeMSLR} 
is based on algebraic domain decomposition and partitions the symmetrized adjacency graph  recursively into several non-overlapping partitions via a $p$-way vertex separator, where $p$ is an integer multiple of the total number of MPI processes. From a numerical perspective, \texttt{parGeMSLR} builds a Schur complement approximate inverse preconditioner as the sum between the matrix inverse of the 
interface coupling matrix and a low-rank correction term. To reduce the cost associated with the computation of the approximate inverse matrices, \texttt{parGeMSLR} exploits a multilevel partitioning of the algebraic domain. The \texttt{parGeMSLR} library is implemented on top of the Message Passing Interface and can solve both real and complex linear systems. Furthermore, 
\texttt{parGeMSLR} can take advantage of hybrid computing environments with in-node access to 
one or more Graphics Processing Units. Finally, the parallel efficiency (weak and strong scaling) 
of \texttt{parGeMSLR} is demonstrated on a few model problems arising from discretizations of 3D Partial Differential Equations.
\end{abstract}

\begin{keyword}
Schur complement, low-rank correction, distributed-memory preconditioner, sparse non-Hermitian linear systems, Graphics Processing Units
\end{keyword}

\maketitle
%---------------------------------------------
% Introduction
%---------------------------------------------

\section{Introduction}
This paper discusses a distributed-memory library for the iterative solution 
of systems of linear algebraic equations of the form
\begin{equation}
  Ax=b,
  \label{eq:Ax=b}
\end{equation}
where the matrix $A\in \mathbb{C}^{n\times n}$ is large, sparse, and 
(non-)Hermitian. Problems of this form typically originate from the 
discretization of a Partial Differential Equation in 2D or 3D domains. 

Iterative methods solve \eqref{eq:Ax=b} by a preconditioned Krylov subspace 
iterative methods \cite{saad_iterative_2003,van2003iterative}, e.g., 
preconditioned Conjugate Gradient \cite{saad_iterative_2003}, if $A$ is Hermitian 
and positive-definite, or GMRES \cite{saad_gmres_1986} if $A$ is non-Hermitian. 
The role of the preconditioner is to cluster the eigenvalues in an effort to 
accelerate the convergence of Krylov subspace method. 
For example, an efficient right preconditioner $M$ transforms 
\eqref{eq:Ax=b} into the preconditioned system $AM^{-1}(Mx)=b$, where 
$M^{-1}$ can be applied inexpensively. An additional requirement is that the setup 
and application of the operator $M^{-1}$ should be easily parallelizable.

Similarly to Krylov subspace methods, algebraic multigrid (AMG) methods are 
another widely-used class of iterative solvers \cite{ruge_algebraic_1987}. 
AMG uses the ideas of interpolation and restriction to build multilevel 
preconditioners that eliminate the smooth error components. AMG is provably 
optimal for Poisson-like problems on regular meshes where the number of 
iterations to achieve convergence almost stays constant as the problem size 
increases. This property leads to appealing weak scaling results of AMG in 
distributed-memory computing environments 
\cite{henson_boomeramg_2002,cleary2000robustness,bell2012exposing}. However, 
AMG can fail when applied either to indefinite problems or irregular meshes. 
It is worth mentioning that AMG can also be used as a preconditioner in the 
context of Krylov subspace methods.

For general sparse linear systems, a well-known class of general-purpose 
preconditioners is that of Incomplete LU (ILU) factorization preconditioners \cite{saad_ilut:_1994,chow1997experimental,saad_iterative_2003}. Here, the 
matrix $A$ is approximately factored as $A\approx LU$ where $L$ is
lower triangular and $U$ is upper triangular, and the preconditioner is 
defined as $M=LU$. Applying $M^{-1}$ then consists of two triangular 
substitutions. ILU preconditioners can be applied 
to a greater selection of 
problems than AMG, including indefinite problems such as discretized
Helmholtz equations \cite{ernst_why_2012,helmhotz}, and their robustness 
can be improved by modified/shifted ILU strategies
\cite{magolu_preconditioning_2000,erlangga_comparison_2006,osei-kuffuor_preconditioning_2010}. 
On the other hand, the scalability of ILU preconditioned Krylov
subspace methods is typically inferior compared to AMG. In particular, even for 
Poisson-like problems, the number of iterations to achieve convergence by 
ILU preconditioned Krylov subspace methods increases with respect to 
the matrix size. Moreover, the sequential nature of triangular substitutions
limit the parallel efficiency of ILU preconditioners implemented 
on distributed-memory systems, and 
recent efforts have been focusing on improving their scalability, e.g., see  \cite{anzt2018parilut,anzt2019parilut,chow_fine-grained_2015}.

The parallel efficiency of ILU preconditioners can be enhanced by 
domain decomposition (DD), where the original problem is decomposed 
into several subdomains which correspond to different blocks of rows of 
the coefficient matrix $A$. The simplest DD-based ILU approach is the 
block-Jacobi ILU preconditioner, where a local ILU is performed on each 
local submatrix. Since this method ignores all of the off-diagonal matrices 
corresponding to inter-domain couplings, its convergence rate tends 
to become slower as the number of
subdomains increases, and several strategies have been proposed to handle the inter-domain
couplings in order to improve the convergence rate. Restricted Additive Schwarz
(RAS) methods expand the local matrix by a certain level to gain a faster
convergence rate at the cost of losing some memory scalability
 \cite{cai_restricted_1999}.  Global factorization ILU methods factorize local
rows corresponding to interior unknowns first, after which a global factorization
of the couplings matrix is applied based on some graph algorithms
\cite{hysom_efficient_1999,karypis_parallel_1997}.  
These methods use partial ILU techniques with dropping
\cite{saad_bilutm_1999,li_parms:_2003}, incomplete triangular solve
\cite{nievinski_parallel_2018}, and low-rank approximation
\cite{dillon_hierarchical_2018} to form the Schur complement system and can be
generalized into multilevel ILU approaches
\cite{saad_bilutm_1999,li_parms:_2003,dillon_hierarchical_2018}.
When the Finite Element method is used and the elements are known, two-level DD methods including BDDC \cite{mandel2003convergence} and FETI-DP \cite{farhat2001feti,heinlein2021combining}, as well as the GenEO preconditioner \cite{spillane2014abstract} are also have been shown to be effective approaches. 
We note that an additional strategy is to combine approximate 
direct factorization techniques with low-rank representation of matrix blocks \cite{amestoy2000mumps}, PasTix \cite{henon2002pastix}, and DDLR \cite{li_low-rank_2017}. 
When the matrix $A$ is SPD, it is possible to reduce the size of the Schur complement 
matrix without introducing any fill-in, e.g., see SpaND \cite{boman2020preconditioner}.

Other preconditioning strategies that can be implemented on distributed-memory 
environments include the (factorized) sparse approximate inverse preconditioners
\cite{benzi_sparse_1998,chow_approximate_1998,janna_block_2010,anzt2018incomplete,grote1997parallel}, polynomial preconditioners \cite{ye_preconditioning_2019}, and rank-structured 
preconditioners
\cite{cai_smash_2018,hackbusch_sparse_1999,hackbusch_sparse_2000,xi_superfast_2014}; 
see also \cite{chen2018distributed} for a distributed-memory hierarchical solver. 
Some of the these techniques can be further compounded with AMG, as “smoothers", or 
ILU-based preconditioners. For example, a combination of SLR \cite{li_schur_2016} 
and polynomial preconditioning is discussed in \cite{ye_preconditioning_2019}.

\subsection{Contributions of this paper}

This paper discusses the implementation of a distributed-memory library, 
termed\footnote{The abbreviation of the library is derived by the complete 
name “parallel Generalized multilevel Schur complement Low-Rank 
preconditioner"} \texttt{parGeMSLR}, for the iterative solution of 
sparse systems of linear algebraic equations in large-scale distributed-memory 
computing environments. 
\texttt{parGeMSLR}\footnote{The source code can be found in 
\href{https://github.com/Hitenze/pargemslr}{https://github.com/Hitenze/pargemslr}} 
is written in C++, and communication among different processor groups is 
achieved by means of the Message Passing Interface standard (MPI). The
\texttt{parGeMSLR} library is based on the Generalized Multilevel 
Schur complement Low-Rank (GeMSLR) algorithm described in \cite{dillon_hierarchical_2018}. GeMSLR applies a multilevel partitioning of the algebraic domain, and the variables associated with each level 
are divided into either interior or interface variables. The multilevel 
structure is built by applying a $p$-way graph partitioner to partition 
the induced subgraph associated with the interface variables of the 
preceding level. Once the multilevel partitioning is completed, GeMSLR 
creates a separate Schur complement approximate inverse at each level. 
Each approximate inverse is the sum of two terms, with the first term 
being an approximate inverse of the interface coupling matrix, and 
the second term being a low-rank correction which aims at bridging the 
gap between the first term and the actual Schur complement matrix inverse 
associated with that level. 
Below, we summarize the main features of the \texttt{parGeMSLR} library:
\begin{enumerate}
    \item \textbf{Scalability}. 
    \texttt{parGeMSLR} extends the capabilities of low-rank-based preconditioners, such as 
    GeMSLR, by recursively partitioning the algebraic domain into levels which have the same 
    number of partitions. In turn, 
    this leads to enhanced scalability when running on distributed-memory environments.
    \item \textbf{Robustness and complex arithmetic}.
    In contrast to ILU preconditioners, the numerical method implemented 
    in \texttt{parGeMSLR} is less sensitive to indefiniteness and can be updated on-the-fly 
    without discarding previous computational efforts. Additionally, \texttt{parGeMSLR} 
    supports complex arithmetic and thus can be utilized to solve complex linear systems 
    such as those originating from the discretization of Helmholtz equations. 
    \item \textbf{Hybrid hardware acceleration}.
    GPU acceleration is supported in several iterative solver libraries aiming 
    to speed-up the application of preconditioners such as AMG or ILU, e.g., 
    \texttt{hypre} \cite{falgout_hypre_2002}, \texttt{PARALUTION} \cite{paralution}, 
    \texttt{ViennaCL} \cite{rupp2016viennacl}, 
    \texttt{HIFLOW} \cite{hiflow3},
    \texttt{PETSc} \cite{balay2001petsc}, and \texttt{Trilinos} \cite{trilinos-website}. A number of direct solver libraries including \texttt{STRUMPACK} \cite{ghysels2017robust,ghysels2016efficient,rouet2016distributed} and SuperLU \_DIST \cite{lidemmel03} also provide GPU support.
    Similarly, \texttt{parGeMSLR} can exploit one or more GPUs by offloading 
    any computation for which the user provides a CUDA interface.  
\end{enumerate}

This paper is organized as follows. Section~\ref{sec:alg} discusses low-rank 
correction preconditioners and provides an algorithmic description of 
\texttt{parGeMSLR}. Section~\ref{sec:dist} provides details on the multilevel reordering 
used by \texttt{parGeMSLR}. Section~\ref{sec:imp} presents in-depth discussion and 
details related to the implementation and parallel performance aspects of 
\texttt{parGeMSLR}. Section~\ref{sec:tests} demonstrates the performance of \texttt{parGeMSLR} 
on distributed-memory environments. 
Finally, our concluding remarks are presented in Section~\ref{sec:conclusion}.

\section{Schur complement approximate inverse preconditioners via low-rank corrections}\label{sec:alg}

This section discussed the main idea behind (multilevel) Schur complement 
preconditioners enhanced by low-rank corrections, e.g., see   \cite{grigori:hal-01017448,li_schur_2016,dillon_hierarchical_2018,xi_algebraic_2016}. 

\subsection{The Schur complement viewpoint}

Let the linear system $Ax=b$ be permuted as 
\begin{equation} \label{eq:A0}
A_0x=P^TAP(P^Tx)=P^Tb, 
\end{equation}
where $P$ is an $n\times n$ permutation matrix such that 
\begin{equation*}
A_0=
\begin{bmatrix}
 B & F \\
 E & C \\
\end{bmatrix}
=
\begin{bmatrix}
B^{(1)} & & & & F^{(1)} \\
& B^{(2)} & & & F^{(2)} \\
& & \ddots & & \vdots \\
& & & B^{(p)} & F^{(p)} \\
E^{(1)} & E^{(2)} &\cdots & E^{(p)} & C \\
\end{bmatrix}, 
\end{equation*}
and the matrices $B^{(i)},\ F^{(i)}$, and $E^{(i)}$ are of 
size 
$d_i\times d_i,\ d_i \times s$, and $s\times d_i$, 
respectively. The matrix $C$ is of size $s\times s$, and the 
matrix partitioning satisfies $d+s=\sum\limits_{j=1}^{j=p}d_j+s_j=n$. Such matrix 
permutations can be computed by partitioning the adjacency graph of the matrix $|A|+\left|A^T\right|$ into 
$p\in \mathbb{N}$ non-overlapping partitions and reordering the unknowns/equations 
such that the variables associated with the $d$ interior nodes across all partitions 
are ordered before the variables associated with the $s$ interface nodes. 

Following the above notation, the linear system in (\ref{eq:A0}) can be written in a block form
\begin{equation}
\begin{bmatrix}
 B & F \\
 E & C \\
\end{bmatrix}
\begin{bmatrix}
 u \\ 
 v \\
\end{bmatrix}
=
\begin{bmatrix}
  f \\ 
  g \\
\end{bmatrix}, 
\label{eq:blockAx=b} 
\end{equation}
where $u,f\in \mathbb{R}^d$ and $v,g\in \mathbb{R}^s$. Once the solution in (\ref{eq:blockAx=b}) is computed, the solution $x$ of the original, non-permuted system of linear algebraic equations
$Ax=b$ can be obtained by the inverse permutation $x = P\begin{bmatrix}
 u \\ 
 v \\
\end{bmatrix}$. Throughout the rest of this section we focus on the solution 
of the system in (\ref{eq:blockAx=b}). 

Following a block-LDU factorization of the matrix $A_0$, the permuted linear 
system in (\ref{eq:A0}) can be written as 
\begin{equation}
\begin{bmatrix}
 I & \\ 
 EB^{-1} & I \\
\end{bmatrix}
\begin{bmatrix}
 B & \\
 & S \\
\end{bmatrix}
\begin{bmatrix}
 I & B^{-1} F \\
 & I 
\end{bmatrix}
\begin{bmatrix}
 u \\ 
 v \\
\end{bmatrix}
 = 
\begin{bmatrix}
  f \\ 
  g \\
\end{bmatrix}, 
\label{eq:blockldu_sec2} 
\end{equation}
where $S=C-EB^{-1}F$ denotes the \emph{Schur complement matrix}. The 
solution of (\ref{eq:blockAx=b}) is then equal to 
\begin{equation*}
\begin{bmatrix}
 u \\ 
 v \\
\end{bmatrix}
=
\begin{bmatrix}
 I & -B^{-1}F \\ 
 & I \\
\end{bmatrix}
\begin{bmatrix}
 B^{-1} & \\
 & S^{-1} \\
\end{bmatrix}
\begin{bmatrix}
 I & \\
 -EB^{-1} & I \\ 
\end{bmatrix}\begin{bmatrix}
  f \\ 
  g \\
\end{bmatrix},
\label{eq:blockldusolve_sec2} 
\end{equation*}
which requires: $a$) the solution of two linear systems with the block-diagonal 
matrix $B$, and $b$) the solution of one linear system with the the Schur 
complement matrix $S$. Note that since the matrix $B$ is block-diagonal, the associated 
linear systems are decoupled into $p$ independent systems of linear algebraic 
equations. Assuming a distributed-memory computing environment with 
$p$ separate processor groups, each system of linear algebraic equations can be 
solved in parallel by means of applying a direct solver locally in each separate 
process. 

In several real-world applications, e.g., those involving the 
discretization of PDEs on three-dimensional domains, solving the systems of 
linear algebraic equations with matrices $B$ and $S$ through a direct solver 
is generally impractical, primarily due to the large computational and memory 
cost associated with forming and factorizing the Schur complement matrix. 
An alternative then is to solve the linear systems with matrices $B$ and $S$ 
inexactly. For example, 
the solution of linear systems with matrix $B$ can be computed approximately by 
replacing its exact LU factorization with an incomplete threshold LU (ILUT) \cite{saad_ilut:_1994}. 
Likewise, the exact Schur complement can be sparsified by discarding entries below 
a certain threshold value or located outside a pre-determined pattern 
\cite{li_parms:_2003,rajamanickam2012shylu}
The approximate factorizations of the matrices $B$ and $S$ can be combined 
to form an approximate LDU factorization of (\ref{eq:blockldu_sec2}) 
which can be then used as a preconditioner in a Krylov subspace iterative 
solver such as GMRES. 

\subsection{Schur complements and low-rank corrections} 

One of the main drawbacks associated with incomplete factorizations is that they 
can not be easily updated if one needs a more accurate preconditioner
{unless the iterative ParILUT \cite{anzt2018parilut,anzt2019parilut} works for the problem and is used}. Moreover, 
their robustness can be limited when the matrix $A$ is indefinite. For such 
scenarios, it has been advocated to add a \emph{low-rank correction term} to enhance 
the efficiency of the Schur complement preconditioner, without discarding the 
previously computed incomplete factorizations. The low-rank enhancement implemented  
in \texttt{parGeMSLR} follows the GeMSLR multilevel preconditioner \cite{dillon_hierarchical_2018}, a non-Hermitian 
extension of \cite{li_schur_2016,xi_algebraic_2016}. Other approaches based on low-rank corrections can be found in  \cite{grigori:hal-01017448,daas2021two}. 

The GeMSLR preconditioner expresses the Schur complement matrix as 
\begin{equation} \label{eq:gemslr1}
    S = (I-EB^{-1}FC^{-1})C = (I-G)C,
\end{equation}
where $G=EB^{-1}FC^{-1}$. Consider now the complex Schur decomposition 
$G = WRW^H$, 
where the $s\times s$ matrix $W$ is unitary and the $s\times s$ matrix $R$ 
is upper-triangular such that its diagonal entries contain the eigenvalues 
of matrix $G$. Plugging the latter in (\ref{eq:gemslr1}) results to 
\begin{equation*}\label{eq:gemslr3}
    S = (I-WRW^H)C = W(I-R)W^HC,
\end{equation*}
from which we can write the inverse of the Schur complement matrix as (Sherman-Morrison-Woodbury formula): 
\begin{equation}\label{eq:gemslr4}
    S^{-1} = C^{-1}+C^{-1}W[(I-R)^{-1}-I]W^H.
\end{equation}
Following (\ref{eq:gemslr4}), a system of linear equations with the Schur 
complement matrix requires the solution of a system of linear equations 
with matrix $C$, as well as matrix-vector multiplications and triangular 
matrix inversions with matrices $W$/$W^H$ and $(I-R)^{-1}$, respectively. 
The product of matrices $W[(I-R)^{-1}-I]W^H$ is a Schur decomposition 
by itself, with corresponding eigenvalues $\gamma_i/(1-\gamma_i),\ 
i=1,\ldots,s$, where $\gamma_i$ denotes the $i$-th eigenvalue of the matrix 
$G$. Therefore, as long as the eigenvalues of the latter matrix are not 
located close to one, the matrix $C(S^{-1}-C^{-1})=W[(I-R)^{-1}-I]W^H$ 
can be approximated by a low-rank matrix, i.e., $S^{-1}$ is approximately 
equal to $C^{-1}$ plus some low-rank correction.

The expression in (\ref{eq:gemslr4}) can be transformed into a practical 
preconditioner if the matrix $W[(I-R)^{-1}-I]W^H$ is replaced by a 
rank-$k$ approximation, where $k\in \mathbb{N}$ is generally a user-given 
parameter. More specifically, let $W_k$ denote the 
$s\times k$ matrix which holds the leading $k$ Schur vectors of matrix 
$G$, and let $R_k$ denote the $k\times k$ leading principal submatrix 
of matrix $R$. Then, the GeMSLR approximate inverse preconditioner is  
equal to
\begin{equation}\label{eq:gemslr6}
    M^{-1} = C^{-1}+C^{-1}W_k[(I-R_k)^{-1}-I]W_k^H \approx S^{-1}.
\end{equation}

\subsection{Computations with an incomplete factorization of $B$}

For large-scale problems, computing an exact factorization of the block-diagonal 
matrix $B$ can be quite expensive. Instead, what is typically available is an 
ILUT factorization $LU\approx B$. Therefore, instead of computing a  
rank-$k$ Schur decomposition of matrix $G$, in practice we approximate a truncated 
Schur decomposition of the matrix $\widehat{G}=E(U^{-1}L^{-1})FC^{-1}$. Let then 
\begin{equation*}
\widehat{G} \widehat{V}_{m} = \widehat{V}_{m}\widehat{H}_{m} + \beta_m \widehat{v}_{m+1}e_m^H, 
\end{equation*}
denote an $m$-length Arnoldi relation obtained 
with matrix $\widehat{G}$, where $[\widehat{V}_{m},\widehat{v}_{m+1}]^H[\widehat{V}_m,\widehat{v}_{m+1}]=I$, and $\widehat{H}_{m}$ is upper-Hessenberg. Moreover, let $\widehat{H}_m=QTQ^H$ denote the complex Schur decomposition of matrix $\widehat{H}_m$. The low-rank correction term used in GeMSLR 
is of the form 
$\widehat{W}_k[(I-\widehat{R}_k)^{-1}-I]\widehat{W}_k^H$, where 
$T_k\in \mathbb{R}^{k\times k}$ denotes the $k\times k$ leading principal submatrix 
of matrix $T$, and $\widehat{W}_k=\widehat{V}_mQ_k$, where 
$Q_k \in \mathbb{R}^{s\times k}$ denotes the matrix holding the $k$ leading Schur 
vectors of matrix $\widehat{H}_m$.

\subsection{Multilevel extensions}

For large-scale, high-dimensional problems, the application of the matrix $C^{-1}$  
by means of an LU factorization of matrix $C$ can still be expensive; especially 
when the value of $p$ is too large, leading to large vertex separators. The 
idea suggested in \cite{xi_algebraic_2016,li_low-rank_2017}, and employed by GeMSLR, 
is to take advantage of the purely algebraic formulation developed in the previous section and apply $C^{-1}$ inexactly by using the Schur complement low-rank preconditioner described in the previous section. In fact, this approach can be repeated more than once, leading to a multilevel preconditioner.

More specifically, let $l_{ev}\in \mathbb{N}$ denote the number of levels, and 
define the sequence of matrices 
\begin{equation}
A_l=P_{l-1} C_{l-1} P_{l-1} = 
\begin{bmatrix}
B_l & F_l\\
E_l & C_l\\
\end{bmatrix},\ \ \ \ C_{-1}=A,\ \ \ \ l=0,1,\ldots,l_{ev}-1,
\end{equation}
where the matrix $B_l$ is block-diagonal with $p$ on-diagonal matrix blocks. 
The $2\times 2$ block matrix partition of each matrix $A_l$ is obtained 
by partitioning the adjacency graph of the matrix $|C_{l-1}|+ |C_{l-1}^T|$ 
into $p$ non-overlapping partitions and reordering the unknowns/equations 
such that the variables associated with the interior nodes across all partitions 
are ordered before the variables associated with the interface nodes of the 
adjacency graph. The matrix $C_{l-1}$ is then permuted in-place through the 
$s_{l-1}\times s_{l-1}$ permutation matrix $P_{l-1}$, where $s_{l-1}$ denotes 
the size of the matrix $C_{l-1}$. 

The solution of a system of linear algebraic equations with matrix $A_l$ as the coefficient matrix 
and $\begin{bmatrix}
  f_l^T &  g_l^T
\end{bmatrix}^T$ as the right-hand side, can be computed as 
\begin{equation*}
\begin{bmatrix}
 u_l \\ 
 v_l \\
\end{bmatrix}
=
\begin{bmatrix}
 I & -B_{l}^{-1}F_{l} \\ 
 & I \\
\end{bmatrix}
\begin{bmatrix}
 B_{l}^{-1} & \\
 & S_{l}^{-1} \\
\end{bmatrix}
\begin{bmatrix}
 I & \\
 -E_{l}B_{l}^{-1} & I \\ 
\end{bmatrix}
\begin{bmatrix}
  f_l \\ 
  g_l \\
\end{bmatrix},
\label{eq:blockldusolve_sec22} 
\end{equation*}
where $S_l = C_l-E_lB_l^{-1}F_l$ denotes the $s_l\times s_l$ Schur complement 
matrix associated with the $l$-th level, where $s_l\in \mathbb{N}$ denotes the 
size of the matrix $C_l$. Instead of computing the exact LU factorizations of 
matrices $B_l$ and $S_l$, the preconditioner implemented in the \texttt{parGeMSLR} library 
substitutes $B_l^{-1}\approx (L_lU_l)^{-1}$, where $L_lU_l$ denotes an ILUT 
factorization of matrix $B_l$, and 
\begin{equation}\label{eq:gemslr5}
    S_l^{-1} \approx C_l^{-1}+C_l^{-1}W_{l,k}[(I-R_{l,k})^{-1}-I]W_{l,k}^H,
\end{equation}
where $\widehat{W}_{l,k}$ denotes the matrix which holds the 
approximate leading $k$ Schur vectors of the matrix $\widehat{G}_l=E^T_lU_l^{-1}L_l^{-1}F_lC_l^{-1}$, and 
$\widehat{R}_{l,k}$ denotes the approximation of the $k\times k$  
leading principal submatrix of the matrix $\widehat{R}_l$ that satisfis the Schur decomposition $\widehat{G}_l=\widehat{W}_l\widehat{R}_l\widehat{W}_l^H$.  Algorithm~\ref{alg:gemslrsetup} 
summarizes the above discussion (“setup phase") in the form of an algorithm. Notice that the recursion stops at level $l_{ev}-1$, and an ILUT of the matrix $C_{l_{ev}-1}$ is computed explicitly. 

\begin{algorithm}
\caption{Parallel GeMSLR Setup}\label{alg:gemslrsetup}
\begin{algorithmic}[1]
\Procedure{pGeMSLRSetup}{$A,l_{ev}$} %\Comment{Setup the parallel GeMSLR preconditioner}
  \State Generate $l_{ev}$-level structure by Algorithm \ref{alg:gemslrreorder}.
  \For {$l$ from 0 to $l_{ev}-1$}
    %\State If $l=l_{ev}-1$, compute an ILUT factorization %$L_{l_{ev}-1}U_{l_{ev}-1}\approx A_{l_{ev}-1}$; exit.
    \State Compute ILU factorization $L_{l}U_{l}\approx B_{l}$.
    \State Compute matrices $\widehat{W}_{l,k}$ and $\widehat{R}_{l,k}$.
    \State If $l=l_{ev}-1$, compute an ILUT factorization $L_{l_{ev}-1}U_{l_{ev}-1}\approx C_{l_{ev}-1}$; exit.
  \EndFor
  %\State \Return
\EndProcedure
\end{algorithmic}
\end{algorithm}

Algorithm~\ref{alg:gemslrsolve} outlines the procedure associated 
with the application of the GeMSLR preconditioner (“solve phase"). 
At each level, the preconditioning step consists of a forward and 
backward substitution with the ILUT triangular factors of $B_l$, 
followed by the application of the rank-$k$ correction term. When 
$l=l_{ev}-1$, there is no low-rank correction term applied, since 
this is the last level. Moreover, when $l=0$ (root level), it is 
possible to enhance the GeMSLR preconditioner by applying a few 
steps of right preconditioned GMRES. Note though that 
these iterations are performed with the inexact Schur complement 
$\widehat{S}_l=C_l-E_l(U^{-1}L^{-1})F_l$.

\begin{algorithm}
\caption{Standard Parallel GeMSLR Solve}\label{alg:gemslrsolve}
\begin{algorithmic}[1]
\Procedure{pGeMSLRSolve}{$b,l$} %\Comment{Solve for $q$ with right-hand-side $b$ at level $l$}
%   \If {$l=l_{ev}-1$}
%     \State Solve $q = U_l^{-1}L_l^{-1}b$.
%   \Else
    \State Apply reordering $\begin{bmatrix}b_1\\b_2\\\end{bmatrix}=P_{l-1}b$.
    \State Solve $z_1 = U_l^{-1}L_l^{-1}b_1$.
    \State Compute $z_2 = b_2 - E_lz_1$. 
    \If {$l=0$}
      \State Solve $\widehat{S}_ly_2=z_2$ by right preconditioned GMRES.
    \Else
      %\State Compute $u_2 = W_{l,k}(R_{l,k}-\theta I)W_{l,k}^Hz_2$. 
      \State Compute $u_2 = \widehat{W}_{l,k}[(I-\widehat{R}_{l,k})^{-1}-I]\widehat{W}_{l,k}^Hz_2$
      \State Call $y_2$ = pGeMSLRSolve($u_2+z_2, l+1$).
    \EndIf
    \State Compute $y_1 = z_1 - U_l^{-1}L_l^{-1}F_ly_2$.
    \State Apply reordering $q=P_{l-1}\begin{bmatrix}y_1\\y_2\\\end{bmatrix}$.
%  \EndIf
  \State \Return $x$
\EndProcedure
\end{algorithmic}
\end{algorithm}

\section{Multilevel reordering}\label{sec:dist}

This section outlines the multilevel reordering approach implemented in the 
\texttt{parGeMSLR} library. For simplicity, we focus on symmetric reorderings 
obtained by applying a $p$-way vertex separator to the adjacency graph associated 
with the matrices $|C_{l-1}|+|C_{l-1}^T|,\ 
l=0,\ldots,l_{ev}-1,\ C_{-1}=A$,  
\cite{karypis_fast_1998,catalyurek_hypergraph-partitioning-based_1999,CHACO,SCOTCH}. 
In particular, given a graph $G=(V,E)$, a $p$-way vertex separator computes a 
separator ${\cal S}\subset V$ and $p$ non-overlapping (disjoint) sets 
$V_1,\ldots,V_p \subset V$ such that $V_1\cup\ldots \cup V_p \cup {\cal S}=V$ and 
there are no edges connecting the sets $V_i$ and $V_j$ when $i\neq j$.

\begin{algorithm}
\caption{Parallel GeMSLR Reordering}\label{alg:gemslrreorder}
\begin{algorithmic}[1]
\Procedure{pGeMSLRReordering}{$A,l_{ev}$} 
  \State Set $C_{-1}\equiv A$.
  \For {$l$ from $0$ to $l_{ev}-1$}
    \State Apply $p$-way partitioning to the graph associated with the matrix $|C_{l-1}|+|C_{l-1}^T|$.
    \State Set $A_{l}=P_{l-1}C_{l-1}P_{l-1}=\begin{bmatrix}B_l & F_l \\ E_l & C_l\\\end{bmatrix}$.
    % \If {$l<l_{ev}-1$} 
    % \State Define $A_{l+1}=C_{l}$.
    % \EndIf
  \EndFor
  \State \Return
\EndProcedure
\end{algorithmic}
\end{algorithm}

\subsection{Hierarchical Interface Decomposition}

The GeMSLR preconditioner relies on a Hierarchical Interface Decomposition (HID)  \cite{henon2006parallel} to reduce the setup cost of the ILU and low-rank 
correction parts associated with the setup phase of the preconditioner. The main 
idea behind HID is to partition the adjacency graph of $|A|+|A^T|$ 
into $2^{l_{ev}}$ partitions via nested dissection with a recursion depth of $l_{ev}$. 
The vertex separators at level $l$ are disjoint with each other since they are 
divided by vertex separators from higher levels. When ordered by levels, the 
global permutation of matrix $A$ will have a block-diagonal structure with $2^{l_{ev}-l}$ 
blocks at level $0\leq l \leq l_{ev}-1$, i.e., the number of diagonal blocks at 
each level reduce by a factor of two. 

\subsection{Multilevel partitioning through $p$-way vertex separators}

In contrast to low-rank correction preconditioners such as MSLR and GeMSLR \cite{xi_algebraic_2016,dillon_hierarchical_2018}, the main goal of 
\texttt{parGeMSLR} is to sustain good parallel efficiency, and thus HID is 
not appropriate.\footnote{Nonetheless, HID is offered in \texttt{parGeMSLR}.} 
Instead, the default approach in \texttt{parGeMSLR} is to partition the 
adjacency graph by a multi-level partitioner where each level 
consists of $p$ partitions and a vertex separator. The latter choice results 
to a \emph{fixed number of $p$ partitions at each level}, and thus load 
balancing is generally much better than that obtained using HID. 

\begin{figure*}[!htb]
  \centering
    \includegraphics[width=.3\linewidth]{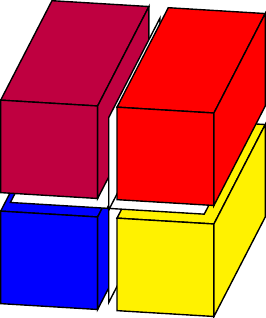}
    \includegraphics[width=.3\linewidth]{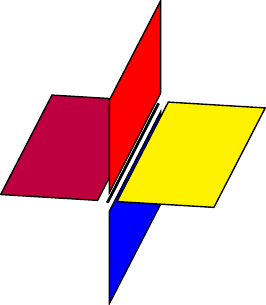}
    \includegraphics[width=.3\linewidth]{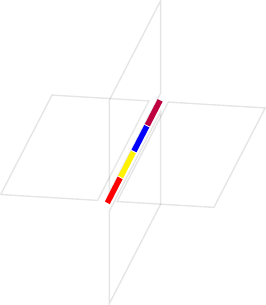}
  \caption{Left: a three-dimensional domain partitioned into $p=4$ subdomains. 
  The vertex separator consists of four faces, with each face located between 
  neighboring subdomains. Center: partitioning of the root-level separator 
  into $p=4$ subdomains. Right: partitioning of the vertex separator at the 
  second level.}
  \label{fig:separator}
\end{figure*}

A high-level description can be found in Algorithm~\ref{alg:gemslrreorder}. 
At the root level ($l=0$), the graph associated with the matrix $|A|+|A^T|$, 
is partitioned into $p$ subdomains with a $p$-way vertex separator, resulting 
to $p$ non-overlapping connected components and their associated vertex 
separator. The multilevel partitioner then proceeds to the next level, $l=1$, 
and applies the $p$-way vertex partitioner to the induced subgraph associated 
with the vertex separator at level $l=0$. This leads to a second set of $p$ non-overlapping connected components and a new, albeit smaller vertex separator. 
The $p$-way vertex partitioner is then applied again to the induced subgraph 
associated with the vertex separator obtained at level $l=1$, etc. The 
procedure continues until either level $l_{ev}-1$ is reached, or the vertex 
separator at the current level $l$ has so few vertices that it can not be further 
partitioned into $p$ non-overlapping partitions. 

An illustration of a three-level, four-way partitioner applied to a three-dimensional 
algebraic domain (a unit cube) is shown in Figure \ref{fig:separator}. The leftmost 
subfigure shows the $p=4$ separate partitions obtained by the application of the four-way vertex partitioner as well as the vertex separator itself (shown in white color) at level $l=0$. This vertex separator, which consists of four two-dimensional 
faces, forms the algebraic object to be partitioned at level $l=1$, and the partitioning is shown in the middle subfigure, where this time the vertex separator 
is a one-dimensional object. Finally, at level $l=2$, the most recent vertex separator is further partitioned into four independent partitions, leading to a new vertex separator which consists of only three vertices; see the rightmost subfigure. 

\begin{figure*}
  \centering
    \includegraphics[width=.45\linewidth]{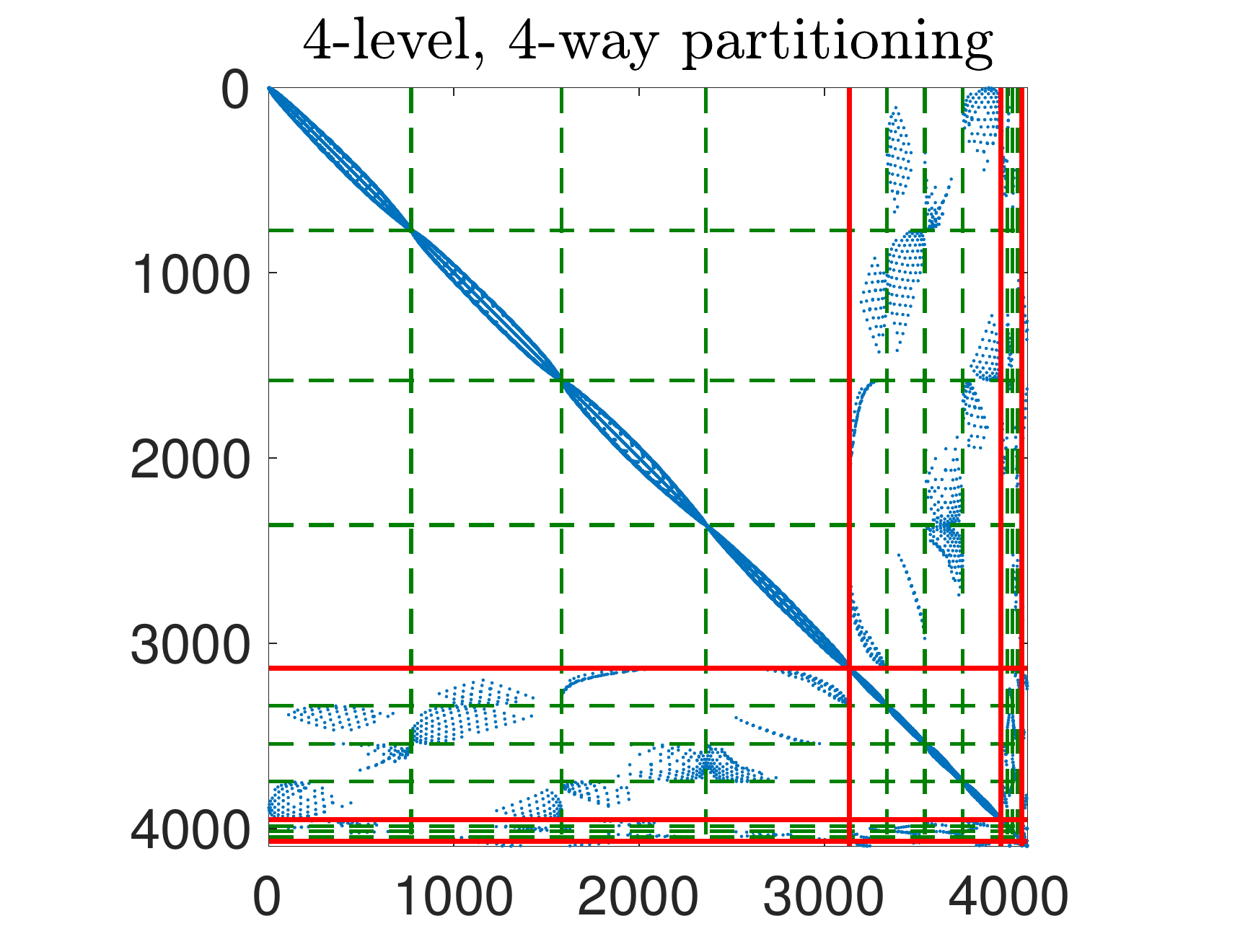}
    \includegraphics[width=.45\linewidth]{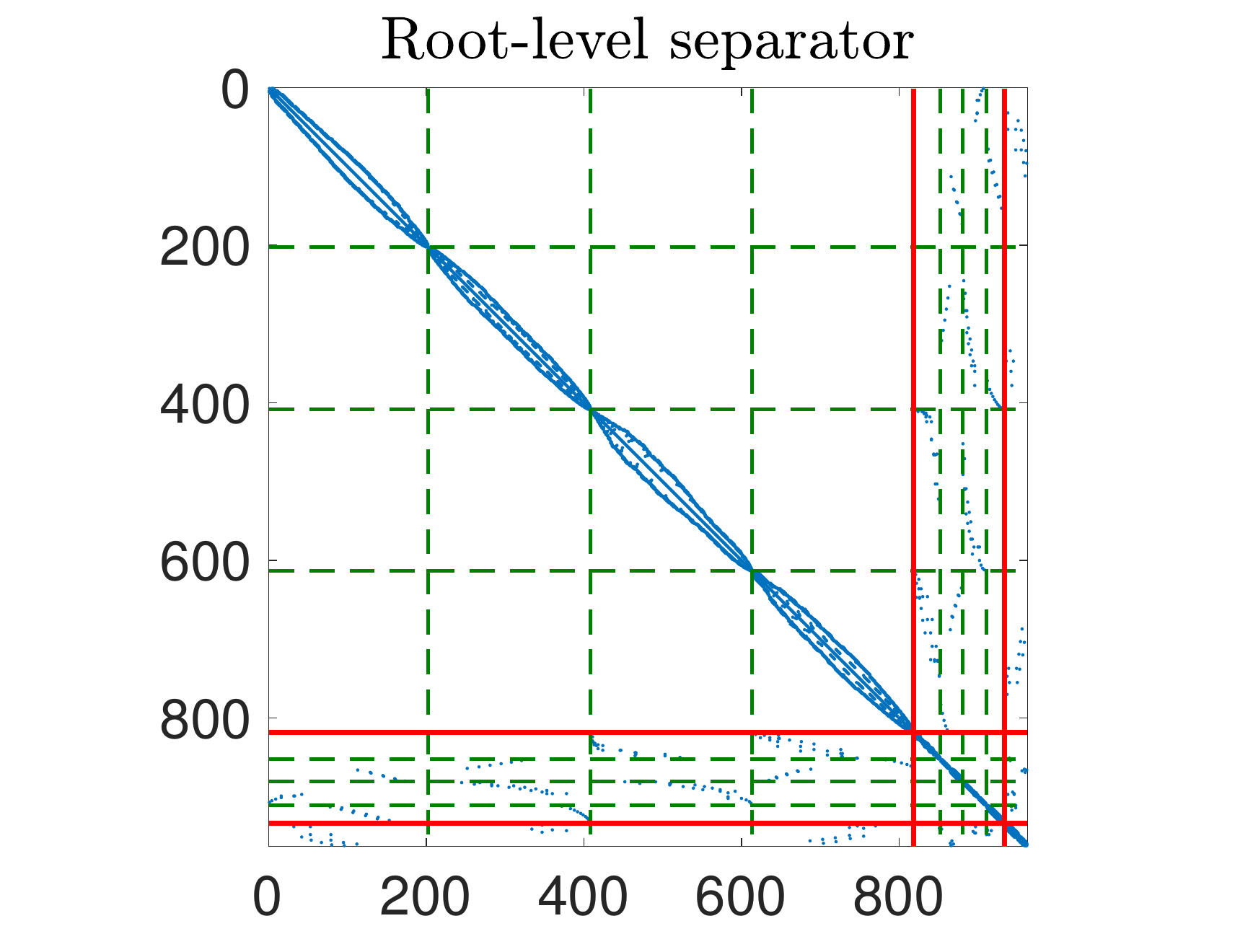}
  \caption{Left: global permutation of matrix $A$ following a multilevel 
  partitioning with 
  $l_{ev}=4$ and $p=4$. Right: zoom-in at the submatrix 
  associated with the permutation of the vertex separators (right-bottom 
  submatrix of the left subfigure).}
  \label{fig:reordering}
\end{figure*}

In addition to the above illustration, Figure \ref{fig:reordering} plots the 
sparsity pattern of a Finite Difference discretization of the Laplace operator 
on a three-dimensional domain, after reordering its rows and columns according 
to a $p$-way, multilevel reordering with $l_{ev}=4$ and $p=4$ (left). A zoom-in 
of the submatrix associated with the permutation of the vertex separators is 
also shown (right). Note that in this particular example, the last level has 
already too few variables to be partitioned any further. In addition to the 
global, multilevel permutation, each matrix $B_l$ can be further permuted 
locally by a reordering scheme such as reverse Cuthill-Mckee (RCM) algorithm or approximate minimal degree algorithm (AMD) \cite{amestoy_approximate_1996,george_computer_1981} to reduce the fill-ins. 

\section{Implementation details of \texttt{parGeMSLR}}\label{sec:imp}

The \texttt{parGeMSLR} library consists of three main modules: 
$a)$ a distributed-memory reordering scheme, $b)$ a Krylov 
subspace iterative accelerator, and $c)$ the 
setup and application of the GeMSLR preconditioner. The 
first module was described in greater detail in Section \ref{sec:dist}, and 
is implemented through a distributed-memory partitioner such as ParMETIS. 
Additional point-to-point communication between neighboring partitions, as well 
as a single All-to-All message are required (to find the new neighbors of each 
partition post-partitioning). Next, we focus on the implementation of the other 
two modules in a distributed-memory environment where different 
processor groups communicate via MPI. 

\subsection{Distributed-memory operations in Krylov accelerators}

Standard, non-preconditioned Krylov iterative methods are built on top of 
simple linear algebraic operations such as matrix-vector multiplication, 
vector scaling and additions, and DOT products. Iterative solvers 
such as GMRES or FGMRES also require the solution of small-scale ordinary 
linear-least squares problems which are typically solved redundantly in 
each MPI process. 

Assuming that the data associated with the system of linear algebraic equations 
we wish to solve is already distributed across the different MPI process via 1D row distribution, AXPY 
operations can be executed locally and involve no communication overhead. On the 
other hand, sparse matrix-vector multiplications and DOT products involve either point-to-point or collective communication. In particular, assume $n_p \in 
\mathbb{N}$ MPI processes. A DOT product then requires a collective operation, i.e., 
{\tt MPI\_Allreduce}, to sum the $n_p$ local DOT products. The cost of this 
operation is roughly $O(log(n_p)\alpha)$, where $\alpha \in \mathbb{R}$ denotes 
the maximum latency between two MPI process. On the other hand, a matrix-vector multiplication with the coefficient 
matrix of the linear system requires point-to-point communication, where the local matrix-vector product in each MPI process consists of operations using local data, 
as well as data associated with MPI processes which are assigned to neighboring 
subdomains, e.g., see \cite{bienz2019node} for additional details and recent advances. 

\subsection{Preconditioner setup and application}

The main module of \texttt{parGeMSLR} is the setup of the GeMSLR preconditioner, followed by 
the application of the latter at each iteration of the Krylov subspace iterative 
solver of choice. Following a multilevel partition into $l_{ev}$ levels (see 
Section \ref{sec:dist}), the setup phase of the GeMSLR preconditioner associated with 
each level $l=0,1,\ldots,l_{ev}-1$, is further divided into two separate submodules: 
$a)$ computation of an ILUT factorization $B_l\approx L_lU_l$, and $b)$ computation 
of an approximate rank-$k$ Schur decomposition of the matrix $\widehat{G}_l=E_l^TU_l^{-1}L_l^{-1}F_lC_l^{-1}$. 

\begin{figure*}[htbp]
  \centering
  \includegraphics[width=0.47\linewidth]{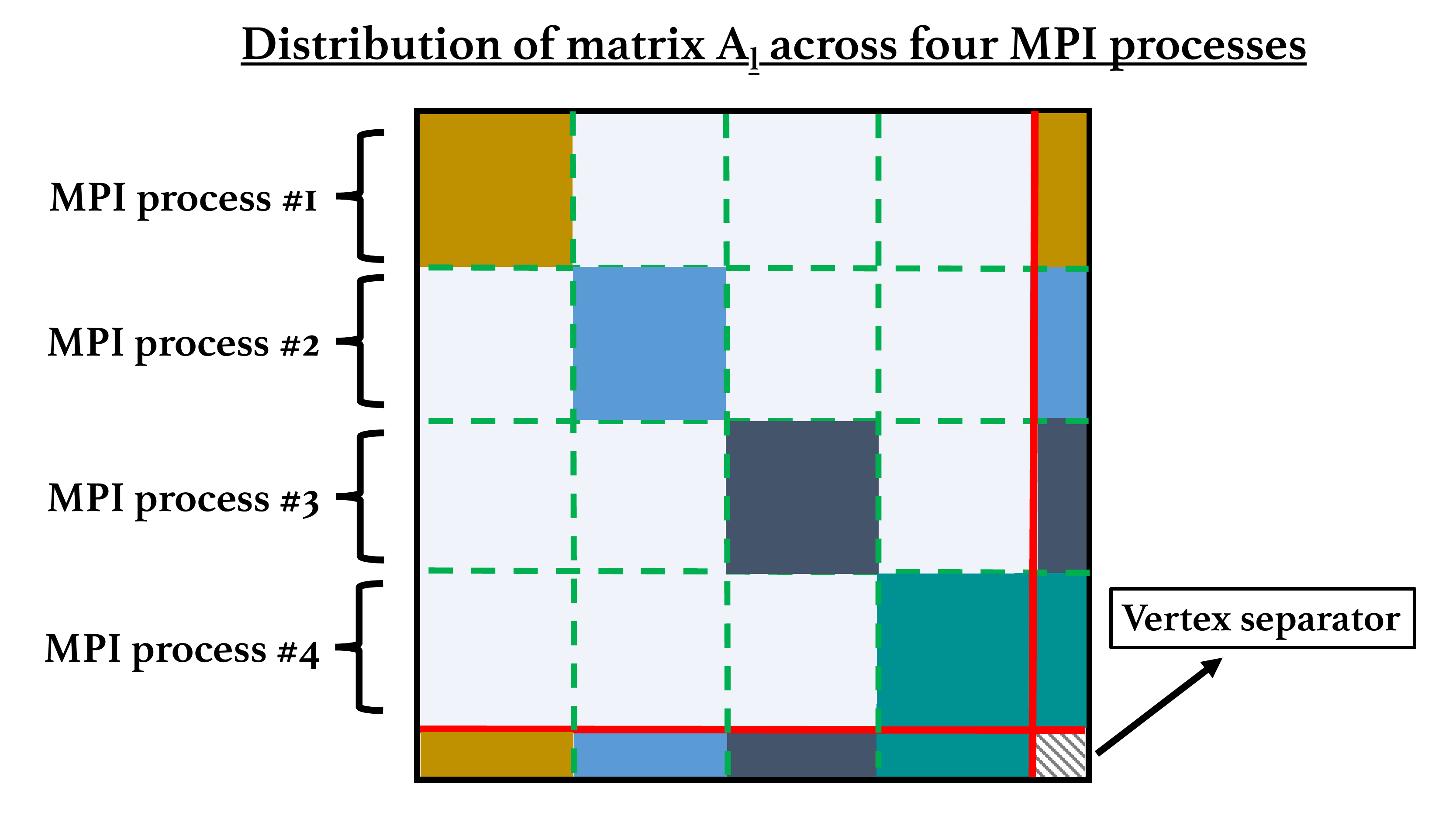}
  \includegraphics[width=0.47\linewidth]{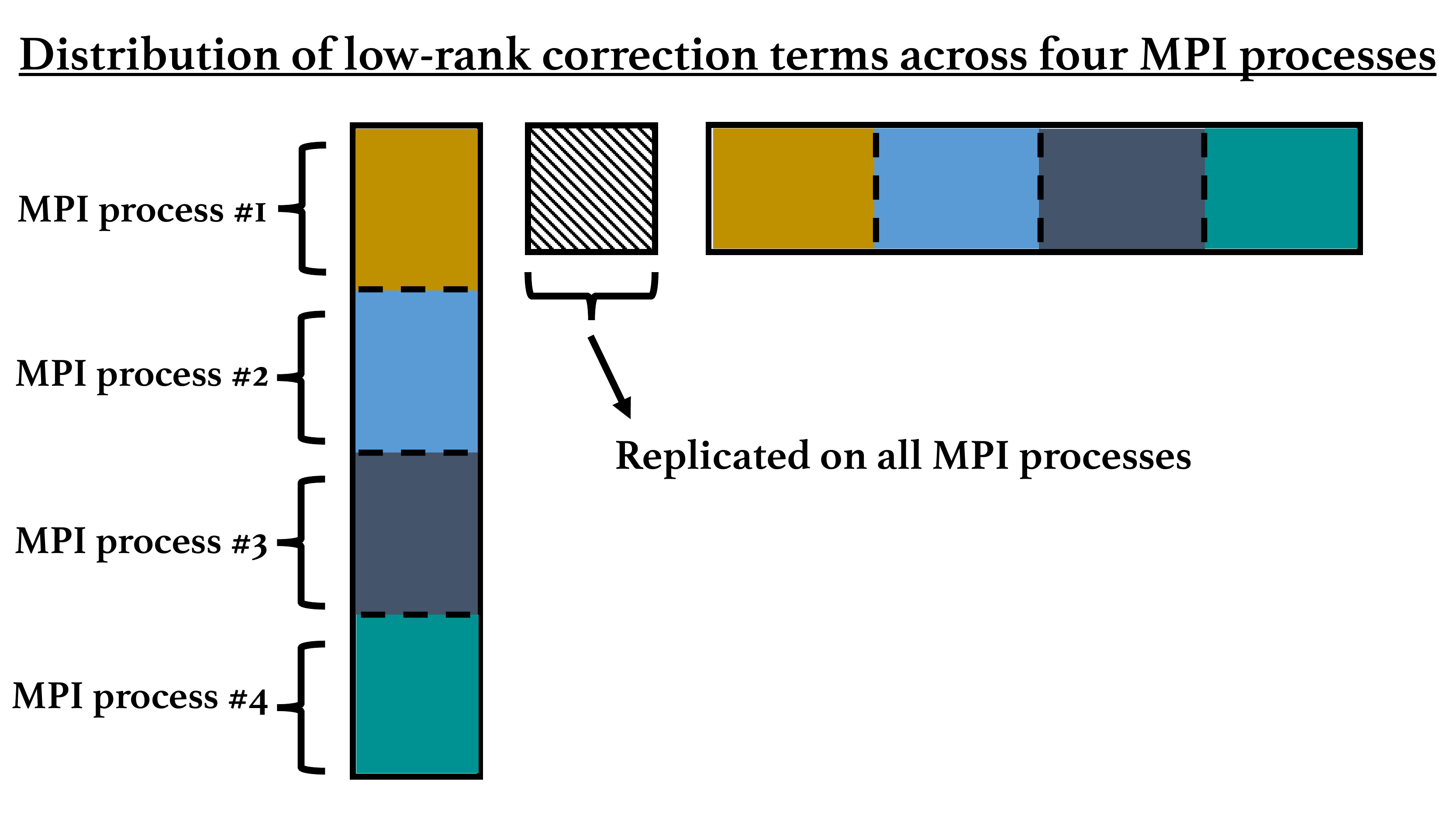}
  \caption{Left: layout of the matrix correction term across four MPI processes (same for any level $0\leq l \leq l_{ev}-1$). Right: layout of a rank-$k$ correction term across four 
  MPI processes (same for any level $0\leq l <l_{ev}-1$).}
  \label{fig:lwrank1}
\end{figure*}

Let us consider each one of the above two tasks separately. Recall that the data 
matrix at each level $0\leq l \leq l_{ev}-1$ has the following pattern
\begin{equation*}
A_l=P_lC_{l-1}P_l=
\begin{bmatrix}
 B_l & F_l \\
 E_l & C_l \\
\end{bmatrix}
=
\begin{bmatrix}
B_l^{(1)} & & & & F_l^{(1)} \\
& B_l^{(2)} & & & F_l^{(2)} \\
& & \ddots & & \vdots \\
& & & B_l^{(p)} & F_l^{(p)} \\
E_l^{(1)} & E_l^{(2)} &\cdots & E_l^{(p)} & C_l \\
\end{bmatrix}. 
\end{equation*}
Now, without loss of generality, assume that each partition is assigned 
to a separate MPI process. Figure \ref{fig:lwrank1} (left) plots a graphical 
illustration of the data layout of matrix $A_l$ obtained by a permutation 
using $p=4$, across four different MPI 
processes. Data associated with separate MPI processes are presented with 
a different color. Notice that the right-bottom 
submatrix denotes the matrix $C_l$ representing the coupling between 
variables of the vertex separator at level $l$. Computing an ILUT 
factorization of the matrix $B_l$ 
decouples into $p$ independent ILUT subproblems $B_l^{(j)}\approx 
L_l^{(j)}U_l^{(j)},\ j=1,\ldots,p$, and thus no communication 
overhead is enabled. On the other hand, the computation of the low-rank correction 
term requires the application of several steps of the Arnoldi iteration, 
and requires communication overhead. 

More specifically, the Arnoldi iteration requires communication among the 
various MPI processes to compute matrix-vector multiplications with the 
iteration matrix 
$\widehat{G}_l$, as well as to maintain orthogonality of the Krylov basis. 
When the latter is achieved by means of standard Gram-Schmidt, Arnoldi 
requires one {\tt MPI\_Allreduce} operation at each iteration. Similarly, 
the matrix-vector multiplication between $\widehat{G}_l$ and a vector $z$ 
is equal to 
\begin{equation*}
\begin{bmatrix}E_l^{(1)} &\cdots & E_l^{(p)}\end{bmatrix}
\begin{bmatrix}
L_l^{(1)}U_l^{(1)} & &  \\
& \ddots &  \\
& & L_l^{(p)}U_l^{(p)}  \\
\end{bmatrix}^{-1}
\begin{bmatrix}
F_l^{(1)}  \\
\vdots   \\
F_l^{(p)} \\
\end{bmatrix}
C_l^{-1}z.
\end{equation*}
The computation of the 
product $C_l^{-1}z$ requires access to the incomplete ILUT factorizations 
and rank-$k$ correction terms associated with all levels $l<\widehat{l}\leq l_{ev}-1$. Therefore, the rank-$k$ correction terms are built in a bottom-up 
fashion, from $l=l_{ev}-1$ to $l=0$, so that level $l$ has immediate access 
to the data associated with all levels $\widehat{l}>l$. 
Once the matrix-vector multiplication $C_l^{-1}z$ is computed, the matrix-vector multiplication with matrix $F_l$ is computed 
with trivial parallelism among the MPI processes, and the same holds for the linear system solutions with matrices $L_l^{(j)},\ U_l^{(j)},\ j=1,\ldots,p$. Finally, the matrix-vector multiplication with matrix $E_l$ requires an {\tt MPI\_Allreduce} 
operation. Note though that if we were to replace vertex separators with edge separators (this option is included in \texttt{parGeMSLR}) then 
the latter multiplication would also be communication-free.

Finally, applying the preconditioner requires embarrassingly parallel 
triangular substitutions with the ILUT factorizations of the 
block-diagonal matrices $B_l$ as well as dense matrix-vector 
multiplications with matrices $\widehat{W}_{l,k},\ \widehat{W}_{l,k}^H$, 
and $(I-\widehat{R}_{l,k})^{-1}$. A matrix-vector multiplication with 
the matrix $\widehat{W}_{l,k}$ requires no communication among the 
MPI processes, while a 
matrix-vector multiplication
with the matrix $\widehat{W}_{l,k}^H$ requires an {\tt MPI\_Allreduce} 
operation at level $l$. Finally, the matrix-vector multiplication 
with the $k\times k$ matrix $(I-\widehat{R}_{l,k})^{-1}$ is performed 
redundantly in each MPI process since $k$ is typically pretty small. 

\subsubsection{Communication overhead analysis}

In this section we focus on the communication overhead associated 
with setting up and applying the preconditioner implemented in 
\texttt{parGeMSLR}. For simplicity, we assume that the number of MPI processes $n_p$ is equal to the number of partitions $p$ at each level. The main parameters of the preconditioner are the number 
of levels $l_{ev}$ and the value of rank $k$.

Let us first consider the application of $m$ Arnoldi iterations 
to compute the matrices $\widehat{W}_{l,k}$ and 
$(I-\widehat{R}_{l,k})^{-1}$ for some $0\leq l \leq l_{ev}-1$. 
As was discussed in the previous section, computing matrix-vector 
products with the matrix $\widehat{G}_l$ requires communication only during the application of the matrices $E_l$ and $C_l^{-1}$. In turn, the latter requires computations with the distributed matrices $C_{l+1}^{-1},\ W_{l+1,k}^H,\ C_{l+2}^{-1},\ W_{l+2,k}^H$, and so on, until we reach level $l_{ev}-1$ where an ILUT of 
the matrix $C_{l_{ev}-1}$ is computed explicitly. Thus, a matrix-vector multiplication 
with the matrix $\widehat{G}_l$ requires $l_{ev}-(l+1)$ (low-rank correction term) and $l_{ev}-l$ ($C_{l}^{-1}$ recursion)  
{\tt MPI\_Allreduce} operations. In summary, 
an $m$-length Arnoldi 
cycle with standard Gram-Schmidt orthonormalization requires 
$(2l_{ev}-2l+1)m$ {\tt MPI\_Allreduce} operations, where we 
also accounted for the two {\tt MPI\_Allreduce} operations 
stemming by Gram-Schmidt and vector normalization at each 
iteration. This communication overhead is inversely proportional to the level index $l$. 
Accounting for all $l_{ev}-1$ levels, the total 
communication overhead associated with the setup phase of the preconditioner 
amounts is bounded by $\delta(k) \sum_{l=0}^{l=l_{ev}-1}(2l_{ev}-2l+1)m$ 
{\tt MPI\_Allreduce} operations, where $\delta(k)\in \mathbb{N}$ denotes 
the maximum number of cycles performed by Arnoldi at any level. 
In \texttt{parGeMSLR}, the default cycle length is $m=2k$ iterations.
Finally, after the set up phase, one full application of the preconditioner 
implemented in the \texttt{parGeMSLR} library requires $2(l_{ev}-l)+1$ {\tt MPI\_Allreduce} 
operations. 

The analysis presented in this section demonstrates that the communication 
overhead associated with the construction of the GeMSLR preconditioner is 
directly proportional to an increase in the value of $l_{ev}$. On the other 
hand, increasing the value of $l_{ev}$ can reduce the computational complexity 
associated with setting up the GeMSLR preconditioner in lower levels. 
Nonetheless, the value of $l_{ev}$ can not be too large, especially when 
the value of $p$ is large, since the size of the vertex separator reduces 
dramatically between successive levels (as is demonstrated in Figure \ref{fig:separator}). 
\subsection{Applying $C_{l_{ev}-1}^{-1}$}

Due to partitioning with a multilevel vertex separator, the matrix 
$C_{l_{ev}-1}$ forms a separate partition which is replicated among 
all MPI processes. Therefore, the simplest approach to apply 
$C_{l_{ev}-1}^{-1}$ is to do so approximately, through computing 
an ILUT redundantly in each MPI process. However, for large problems, 
this approach can quickly become impractical, even if a shared-memory 
variant of ILUT is considered \cite{anzt2018parilut}. On the other 
hand, applying a distributed-memory approach that requires communication 
among the MPI processes can lead to high communication overhead 
since the application of $C_{l_{ev}-1}^{-1}$ is the most common operation 
during the setup phase of the preconditioner. 

\texttt{parGeMSLR} includes several\footnote{See section 2.1 in \url{https://github.com/Hitenze/pargemslr/blob/main/ParGeMSLR/DOCS/Documentation.pdf}} options to apply an approximation of $C_{l_{ev}-1}^{-1}$. The 
default option considered throughout our experiments is to apply 
$C_{l_{ev}-1}^{-1}$ approximately through 
a block-Jacobi approach where $C_{l_{ev}-1}$ is first permuted by 
reverse RCM and then replaced by its on-diagonal block submatrices 
while the rest of the entries are discarded. Generally speaking, 
dropping these entries of $C_{l_{ev}-1}^{-1}$ has minor effects 
since $C_{l_{ev}-1}$ is already close to being block-diagonal for 
modest values of $l_{ev}$ (e.g., three or four) as was already 
demonstrated in Figure \ref{fig:separator}. By default, the number 
of retained on-diagonal blocks of matrix $C_{l_{ev}-1}$ is set equal 
to $p$. The approximate application of $C_{l_{ev}-1}^{-1}$ is then 
trivially parallel among the MPI processes, and each one of the 
retained on-diagonal blocks is applied through ILUT. 

\section{Numerical Experiments}\label{sec:tests}

In this section we demonstrate the parallel performance of \texttt{parGeMSLR}. 
We run our experiments on the \textsc{Quartz} cluster of Lawrence Livermore 
National Laboratory. Each node of \textsc{Quartz} has 128 GB memory and consists 
of 2 Intel Xeon E5-2695 CPUs with 36 cores in total. We use MVAPICH2 2.2.3, to compile \texttt{parGeMSLR} is compiled with MVAPICH2 2.2.3,  following rank-to-core binding. By default, all of the experiments 
presented below are executed in double-precision.\footnote{We note though that 
\texttt{parGeMSLR} supports both real and complex arithmetic, as well as both 
single and double precision.} On top of distributed-memory parallelism, \texttt{parGeMSLR} 
can take advantage of shared memory parallelism using either OpenMP 
or CUDA. The current version of \texttt{parGeMSLR} uses LAPACK for sequential 
matrix decompositions and ParMETIS for distributed graph partitioning \cite{karypis_fast_1998}. 
A detailed documentation of \texttt{parGeMSLR} can be found in the “DOCS" directory of \url{https://github.com/Hitenze/pargemslr}. This documentation provides detailed 
information on how to compile and run \texttt{parGeMSLR}, and includes a detailed 
description of all command-line parameters as well as visualization of the source code hierarchy. Several test drivers, and a sample input file, are also included.

Throughout the rest of this section, we choose Flexible GMRES (FGMRES) with a fixed 
restart size of fifty as the outer iterative solver. The motivation for using FGMRES 
instead of GMRES is that the application of the preconditioner is subject to variations 
due to the application of the inner solver in step 9 of Algorithm~\ref{alg:gemslrsolve}. 
The stopping tolerance for the relative residual norm in FGMRES is set equal to 
$1.0{\tt e}-6$. 
Unless mentioned otherwise, the solution of the linear system $Ax=b$ will be equal to the vector of all ones with an initial approximation equal to zero.
The low-rank correction term at each level consists of approximate Schur vectors 
such that the corresponding approximate eigenvalues are accurate to two digits of 
accuracy, and the restart cycle of thick-restart Arnoldi is equal to $2k$. 

Our distributed-memory experiments focus on the parallel efficiency of \texttt{parGeMSLR} 
both when the problem size remains fixed and $n_p$ increases (strong scaling) and 
the problem size increases at the same rate with $n_p$. In the case of weak 
scaling, the parallel efficiency is equal to $\frac{T_1}{T_{n_p}}$, where $T_1$ and $T_{n_p}$ 
denote the wall-clock time achieved by the sequential and distributed-memory 
version (using $n_p$ MPI processes) of \texttt{parGeMSLR}, respectively. Likewise, in the 
case of strong scaling, the parallel efficiency is equal to $\frac{T_1}{n_pT_{n_p}}$. 
In addition, we also compare 
\texttt{parGeMSLR} against: $a$) the BoomerAMG parallel implementation 
of the algebraic multigrid method in \texttt{hypre}, and 
$b$) the two-level SchurILU approach in \cite{li_parms:_2003}. The latter 
preconditioner uses partial ILU to form an approximation of the Schur 
complement matrix. The preconditioning step is then performed by applying 
GMRES with block-Jacobi preconditioning to solve the linear system associated 
with the sparsified Schur complement. The block-Jacobi preconditioner is applied 
through one step of ILUT, and our implementation of SchurILU is based on the parallel 
ILU(T) in hypre.

Throughout the rest of this section, we adopt the following notation:
\begin{itemize}
\itemsep-0.2em 
    \item $\mathbf{n_p}\in \mathbb{N}$: total number of MPI processes.
    \item $\mathbf{fill}\in \mathbb{R}$: ratio between the number 
    of non-zero entries of 
    the preconditioner and that of matrix $A$.
    \item {\bf p-t} $\in \mathbb{R}$: preconditioner setup time. This 
    includes the time required to compute the 
    ILUT factorizations and low-rank correction terms in \texttt{parGeMSLR}.
    \item {\bf i-t} $\in \mathbb{R}$: iteration time of FGMRES.
    \item $\mathbf{its}\in \mathbb{N}$: total number of FGMRES iterations.
    \item $\mathbf{k}\in \mathbb{N}$: number of low-rank correction terms at each level.
    \item $\mathbf{F}$: flag signaling that FGMRES failed to converge within 1000 iterations.
\end{itemize}

\subsection{A Model Problem}\label{sec:p1}

This section considers a Finite Difference discretization of the model problem
\begin{eqnarray}
\label{eq:test_problem_1}
%-\Delta u -\mathbf{b}\cdot\nabla u -cu &=& f \ \ \text{in} \ \Omega, \nonumber \\
-\Delta u - cu &=& f \ \ \text{in} \ \Omega, \nonumber \\
u &=& 0 \ \ \text{on} \ \partial \Omega.
%,
\end{eqnarray}
We consider a 7-pt stencil and set $\Omega = (0,1)^3$.

\subsubsection{Weak scaling}

Our first set of experiments studies the weak scaling efficiency of \texttt{parGeMSLR}. 
Since varying the values of $l_{ev}$ and $k$ lead to different convergence rates, we  
first consider the case where the number of FGMRES iterations is set equal to thirty, 
regardless of whether convergence was achieved or not. The problem size on each MPI 
process is fixed to $50^3$, while the number of subdomains at each level is set equal 
to $8\times n_p$. Moreover, the number of levels is varied as $l_{ev}\in \{2,3\}$ while 
the rank of the low-rank correction terms is varied as $k\in \{0,100,200\}$. 

%\begin{figure}[htbp]
\begin{figure}[ht]
  \centering
  \begin{subfigure}[b]{0.95\linewidth}
         \centering
         \includegraphics[width=1.\linewidth]{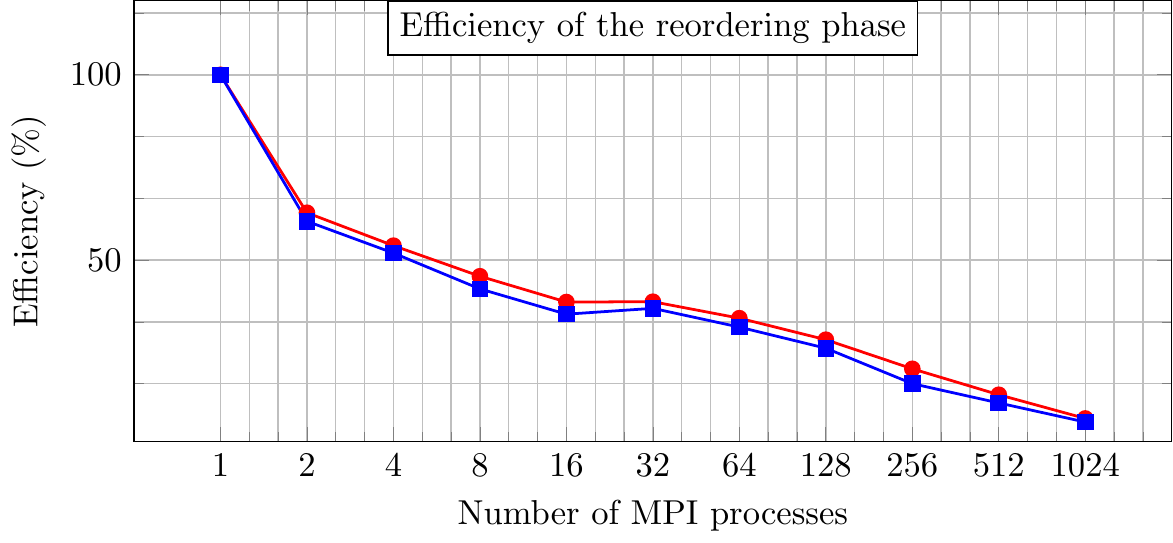}
         %\caption{Efficiency of the reordering phase.}
  \end{subfigure}
  \begin{subfigure}[b]{0.95\linewidth}
         \centering
         \includegraphics[width=1.\linewidth]{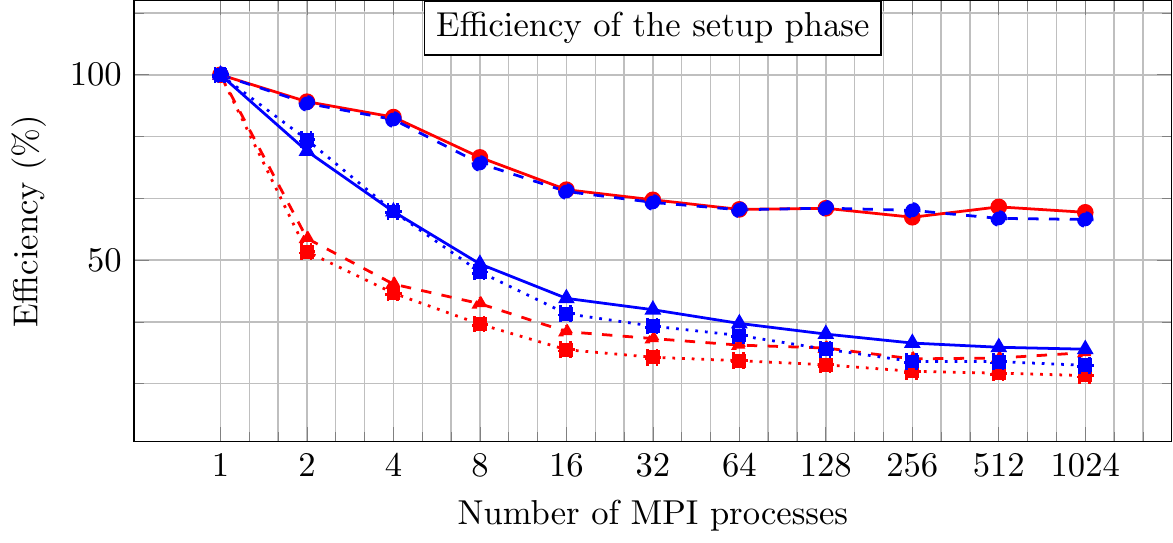}
         %\caption{Efficiency of the setup phase.}
  \end{subfigure}
  \begin{subfigure}[b]{0.95\linewidth}
         \centering
         \includegraphics[width=1.\linewidth]{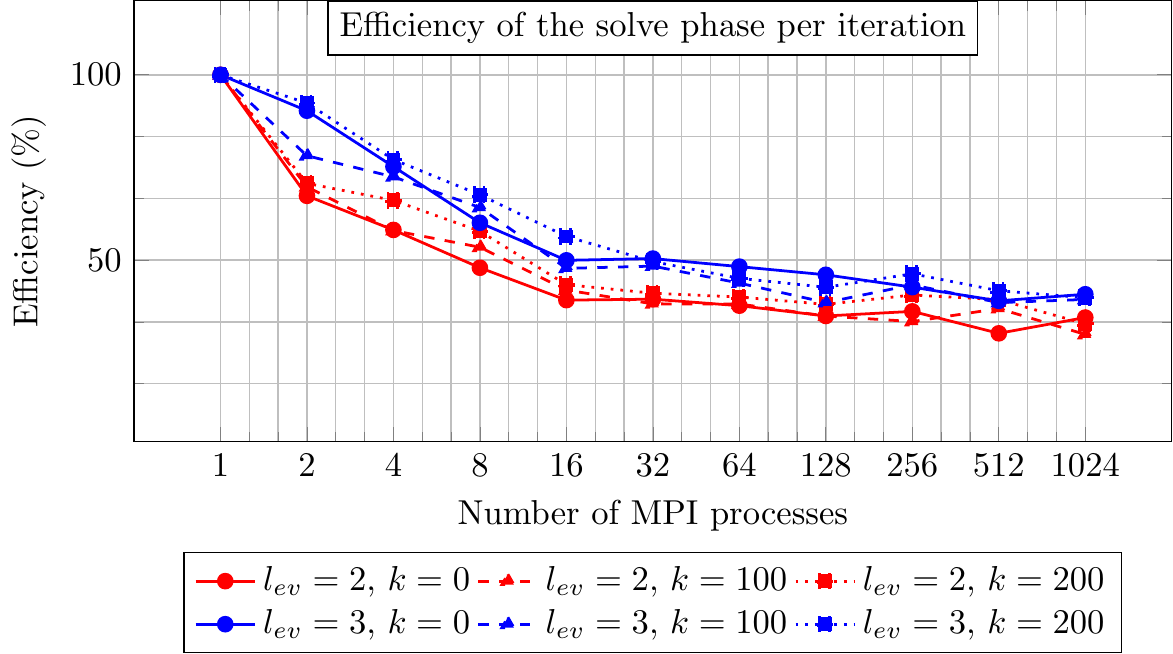}
         %\caption{Efficiency of the solve phase per iteration.}
  \end{subfigure}
  \caption{Weak scaling of \texttt{parGeMSLR} for the Poisson problem when the number of iterations performed by FGMRES is fixed to thirty, and the number of levels is set to 
  $l_{ev}=2$ and $l_{ev}=3$. The number of unknowns on each MPI process is $125,000$, for a maximum problem size $n=800\times400\times400$.}
  \label{fig:ws_pi}
\end{figure}

Figure~\ref{fig:ws_pi} plots the weak scaling efficiency of \texttt{parGeMSLR} 
on up to $n_p=1,024$ MPI processes. The achieved efficiency is similar for both 
options of $l_{ev}$ with a slightly higher efficiency observed for the case 
$l_{ev}=3$. As expected, the highest efficiency achieved during the preconditioner 
setup phase was for the case $k=0$, since there is no communication overhead 
stemming from the low-rank correction terms. Nonetheless, even in this case 
there is some loss in efficiency due to load imbalancing introduced by the ILUT 
factorizations at different levels. Regardless of the value of $k$, the efficiency of 
\texttt{parGeMSLR} drops the most when the number of MPI processes is small, 
regardless of the value of $l_{ev}$. This reduction is owed to the relatively 
large increase on the size of the local Schur complement versus when a larger number 
of MPI processes is utilized. Note though, although not reported in our experiments, 
that the weak scaling efficiency is typically much higher when each MPI process 
handles exactly one subdomain. Finally, the efficiency of the reordering phase 
is rather limited, since the wall-clock time requires to partition the graph 
associated with the matrix $|A|+|A^T|$ and permute the distributed matrix $A$ 
increases as the problem size grows. 

\begin{figure*}
  \centering
  \includegraphics[width=.95\linewidth]{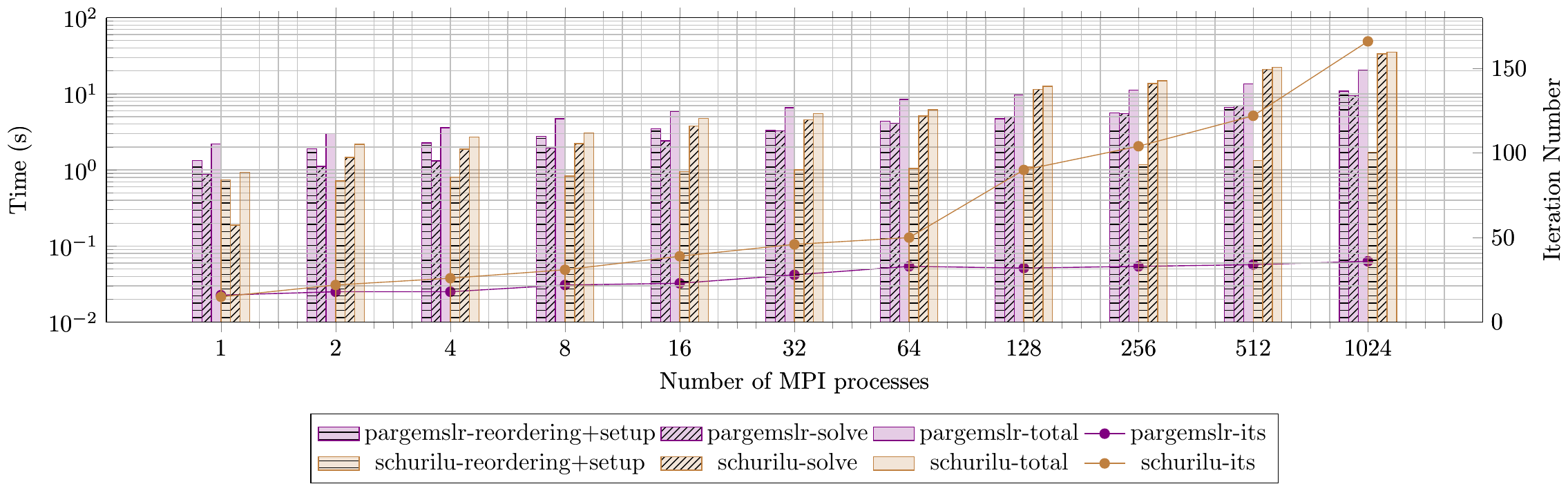}
  \caption{Weak scaling of \texttt{parGeMSLR} and SchurILU on Poisson problems. The 
  number of unknowns on each MPI process is $125,000$, for a maximum problem 
  size $n=800\times400\times400$.}
  \label{fig:ws_1}
\end{figure*}

Figure~\ref{fig:ws_1} plots the weak scalability of \texttt{parGeMSLR} and two-level 
SchurILU, where this time we allow enough iterations in FGMRES until 
convergence. As previously, we use eight subdomains per MPI process, but 
this time we fix $l_{ev}=3$ and $k=10$. In summary, \texttt{parGeMSLR} 
is both faster and more scalable than SchurILU during the solve phase. Moreover, 
\texttt{parGeMSLR} also converges much faster than SchurILU, and the number of total 
FGMRES iterations increases only marginally with the problem size. On the 
other hand, the weak scaling of the preconditioner setup phase of \texttt{parGeMSLR} 
is impacted negatively as the problem size increases due to the need to 
perform more Arnoldi iterations to compute the low-rank correction terms. 

\subsubsection{Strong scaling}

We now present strong scaling results obtained by solving \eqref{eq:test_problem_1} with \texttt{parGeMSLR} on a regular mesh of fixed size as the numbers of MPI processes varies. 
More specifically, the size of the problem is fixed to $n=320^3$ while the number of MPI 
processes varies up to $n_p=1,024$. The values of $l_{ev}$ and $k$ are varied as previously.

\begin{figure}[!htb]
  \centering
  \begin{subfigure}[b]{0.95\linewidth}
         \centering
         \includegraphics[width=1.\linewidth]{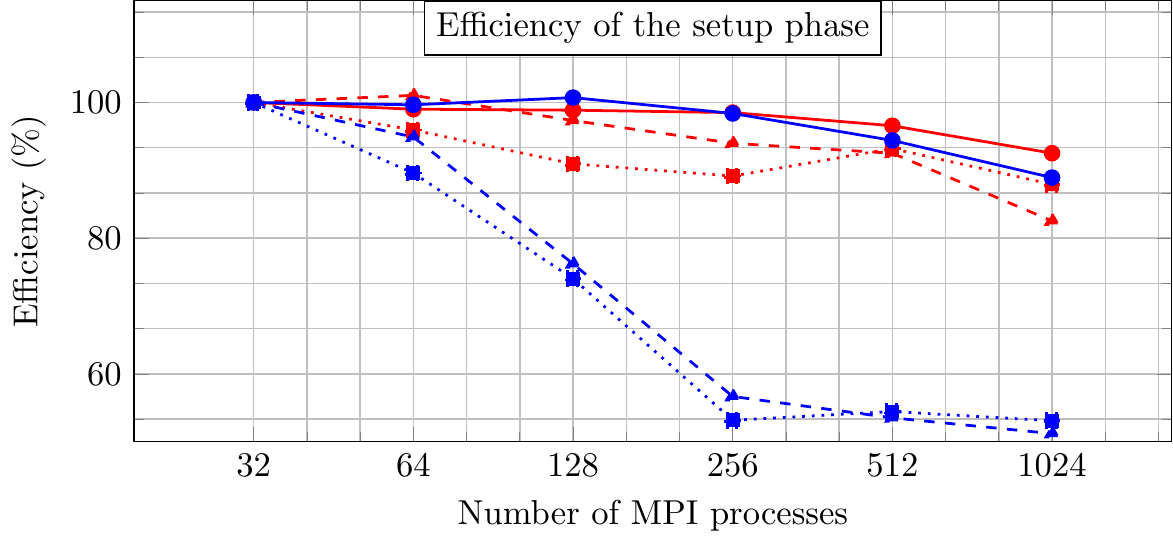}
  \end{subfigure}
  \begin{subfigure}[b]{0.95\linewidth}
         \centering
         \includegraphics[width=1.\linewidth]{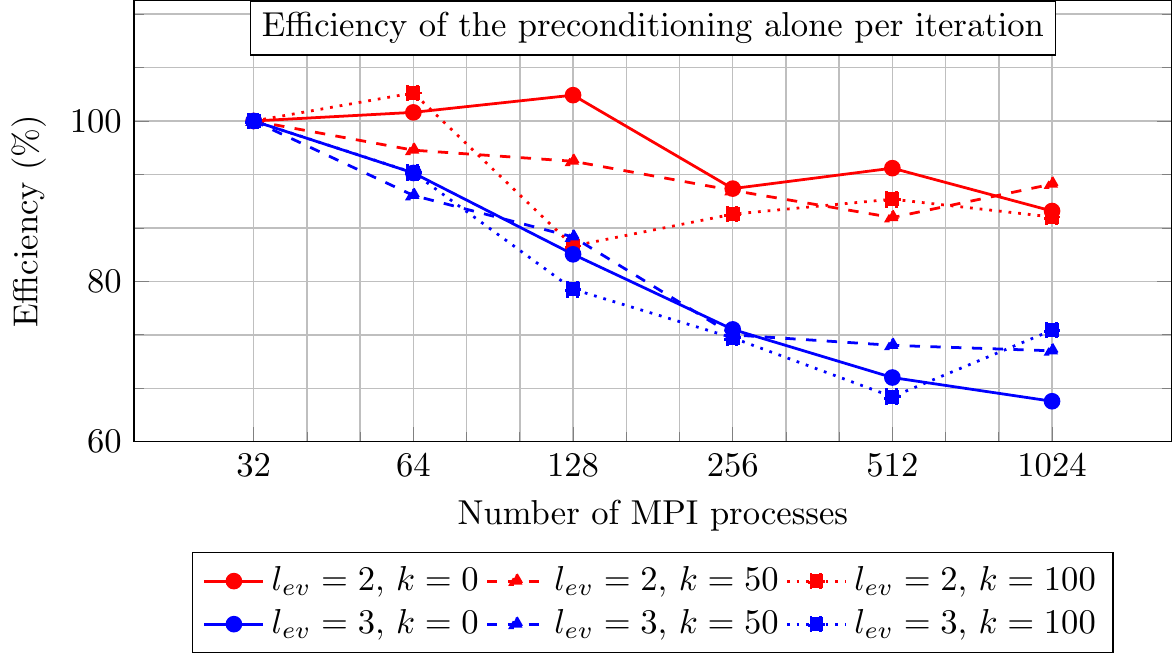}
  \end{subfigure}
  \caption{Strong scaling results for Poisson problems of size $n=320^3$. 
  The number of subdomains is set equal to 2048 in all levels.}
  \label{fig:ss_pi_large}
\end{figure}

Figure~\ref{fig:ss_pi_large} plots the strong scaling of \texttt{parGeMSLR}. In contrast 
to the weak scaling case, setting $l_{ev}=2$ leads to higher efficiency during both 
the setup and application phases of the preconditioner. The reason for this behavior is 
twofold. First, increasing the value of $l_{ev}$ generally deteriorates the effectiveness 
of the preconditioner unless $k$ is large and the threshold used in the local ILUT 
factorizations is small. Second, decreasing the value of $l_{ev}$ enhances strong scalability 
since it leads to smaller communication overheads (i.e., recall the discussion in Section 
\ref{sec:imp}). As a general remark, we note that the setup phase of \texttt{parGeMSLR} generally becomes more expensive in terms of floating-point arithmetic operations as $l_{ev}$ decreases, thus although scalability deteriorates as $l_{ev}$ increases, the actual wall-clock 
time might actually decrease if the number of MPI processes used is small.

\subsection{General Problems}\label{sec:p2}

This section discusses the performance of \texttt{parGeMSLR} on a 
variety of problems in engineering.

\subsubsection{Unstructured Poisson problem on a crooked pipe}

\begin{figure}[htbp]
  \centering
  \includegraphics[width=.45\linewidth]{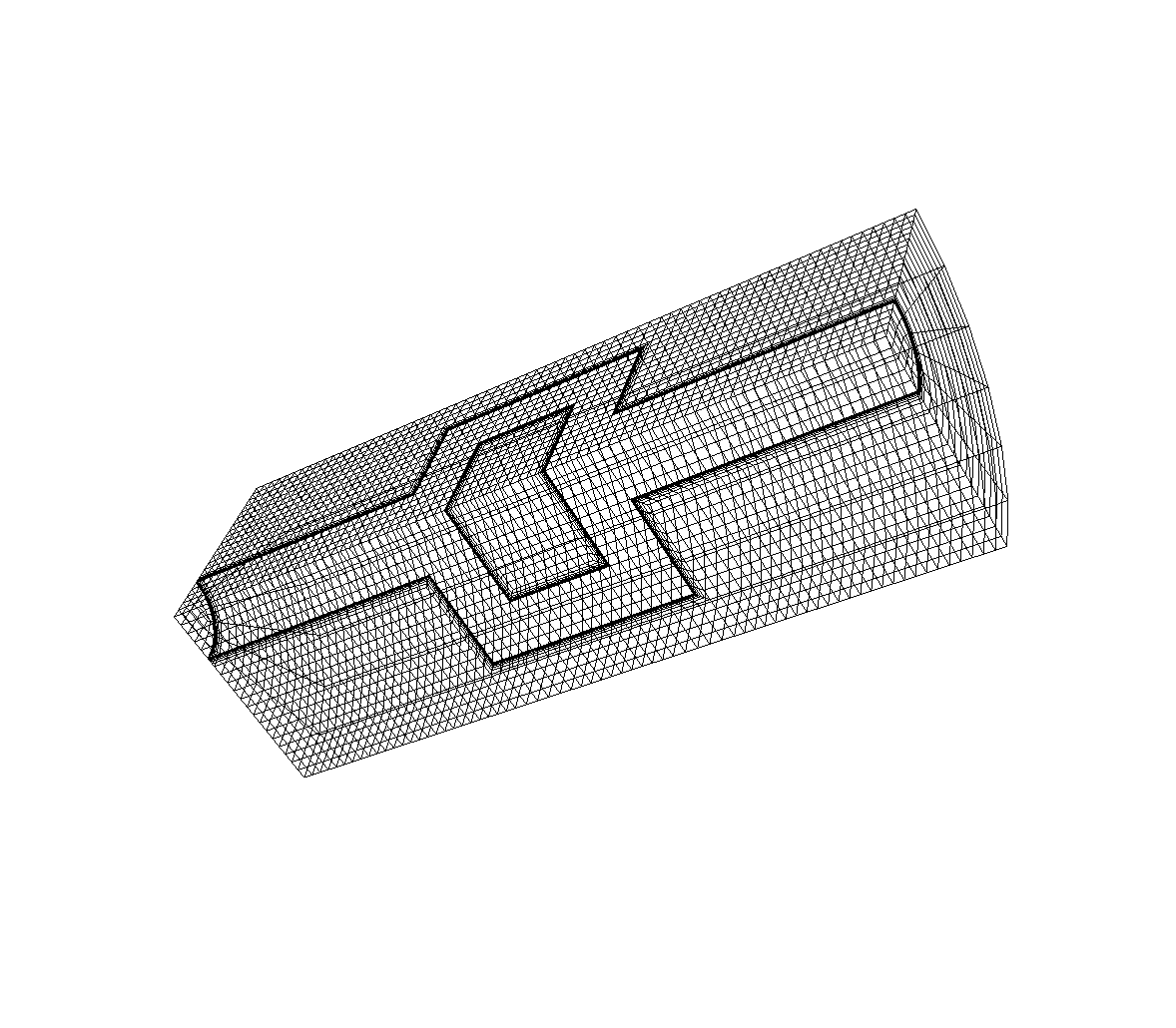}
  \includegraphics[width=.45\linewidth]{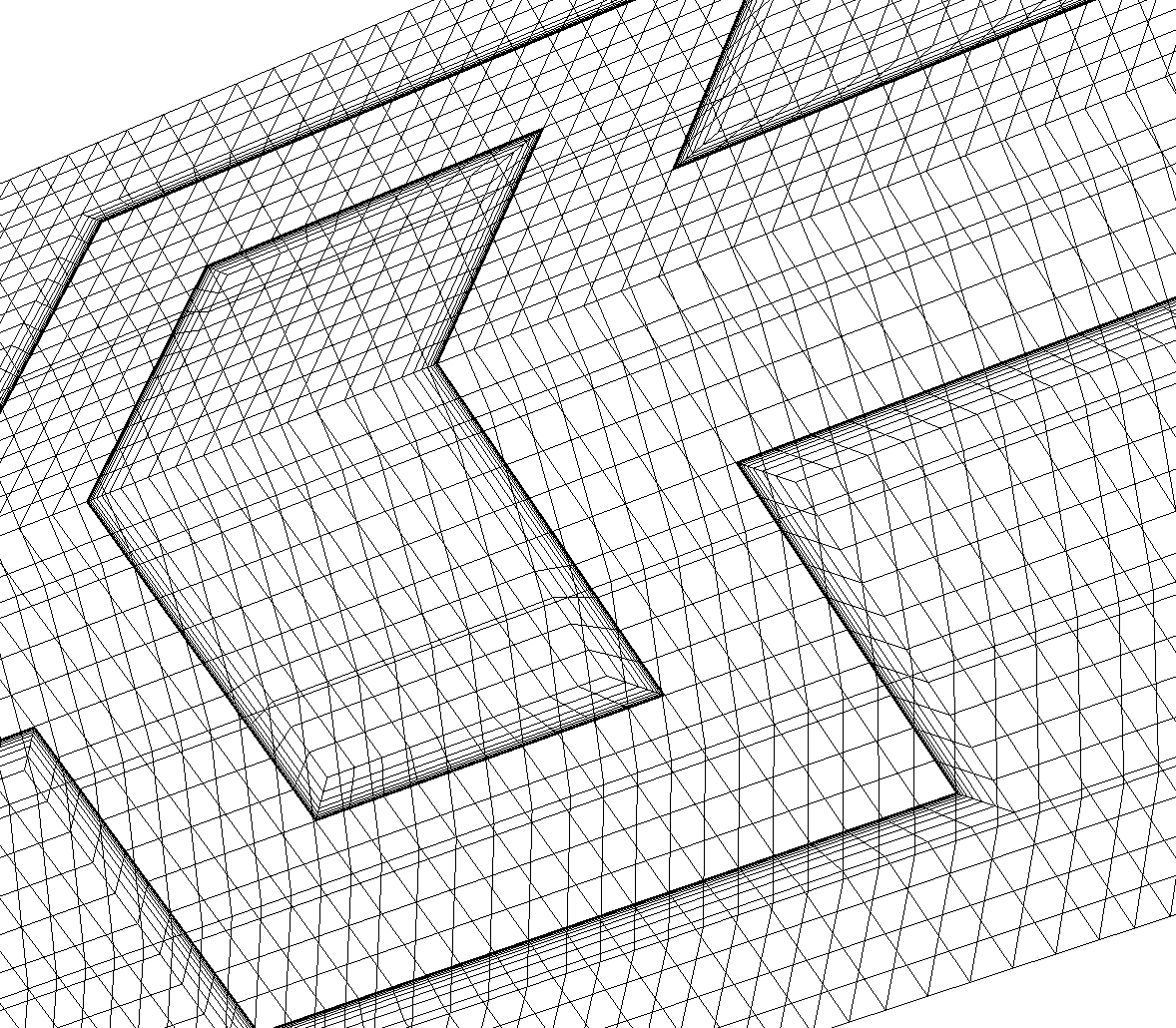}
  \caption{Left: Poisson problem on a crooked pipe mesh. Right: zoom-in of the center part of the mesh.}
  \label{fig:mesh_problem_1}
\end{figure}

We consider the numerical solution of \eqref{eq:test_problem_1} where $f=1$ and $c=0$ on a 3D 
crooked pipe mesh. The problem is discretized by second-order Finite 
Elements using the MFEM library \cite{mfem,mfem-web} with local uniform 
and parallel mesh refinement. 
The initial approximation of the solution is set equal to zero.
We visualize the (inhomogeneous) mesh using 
the package GLVis \cite{glvis-tool} in Figure~\ref{fig:mesh_problem_1}. 
Our experiments consider different refinement levels to generate problems 
of different sizes. Moreover, the maximum number of inner iterations in 
step 9 of Algorithm~\ref{alg:gemslrsolve} is varied between three and five. 
We compare \texttt{parGeMSLR} against BoomerAMG with Hybrid Modified Independent Set 
(HMIS) coarsening, where we consider both Gauss-Seidel and $l_1$ Jacobi 
smoother \cite{baker_multigrid_2011}, and report the corresponding results 
in Table~\ref{tab:test1_crooked_pipe}. \texttt{parGeMSLR} is able to outperform 
Schur ILU, especially for larger problems. Moreover, the iteration time of 
\texttt{parGeMSLR} is similar to that of BoomerAMG with Gauss-Seidel smoother, but much lower than that of BoomerAMG with $l_1$ Jacobi smoother.

\begin{table}[htbp]
\centering
  \caption{Solving (\ref{eq:test_problem_1}) on a crooked pipe mesh.}
  \label{tab:test1_crooked_pipe}
  \begin{tabular}{c|cc|ccccc}
    \hline
     prec & size & $n_p$ & k & fill & p-t & i-t & its \\
  \hline
   \multirow{3}{*}{\shortstack{Boomer\\AMG\\GS}} 
    & 126,805 & 16 & 
    - & 1.71 & 0.17 & 0.69 & 106 \\
    & 966,609 & 32 & 
    - & 1.79 & 0.79 & 5.7 & 198 \\ 
    & 7,544,257 & 64 & 
    - & 1.81 & 3.36 & 45.12 & 250 \\
   \hline
   \multirow{3}{*}{\shortstack{Boomer\\AMG\\Jacobi}} 
    & 126,805 & 16 & 
    - & 1.71 & 0.18 & 1.29 & 226 \\
    & 966,609 & 32 & 
    - & 1.79 & 0.8 & 10.95 & 431 \\ 
    & 7,544,257 & 64 & 
    - & 1.81 & 3.39 & 72.1 & 568 \\
   \hline
   \multirow{3}{*}{\shortstack{Schur\\ILU}} 
    & 126,805 & 16 & 
    - & 1.53 & 0.22 & 0.51 & 65 \\
    & 966,609 & 32 & 
    - & 1.86 & 1.2 & 12.46 & 383 \\
    & 7,544,257 & 64 & 
    - & 1.94 & 5.51 & - & F \\
   \hline
   \multirow{3}{*}{\shortstack{par\\GeMSLR}} 
    & 126,805 & 16 & 
    10 & 1.05 & 0.54 & 0.46 & 25 \\
    & 966,609 & 32 & 
    10 & 1.18 & 3.59 & 4.70 & 53 \\
    & 7,544,257 & 64 & 
    10 & 1.32 & 11.76 & 48.35 & 128 \\
   \hline
  \end{tabular}
\end{table}

\subsubsection{Linear elasticity equation}

\begin{figure}[ht]
  \centering
  \includegraphics[width=.75\linewidth]{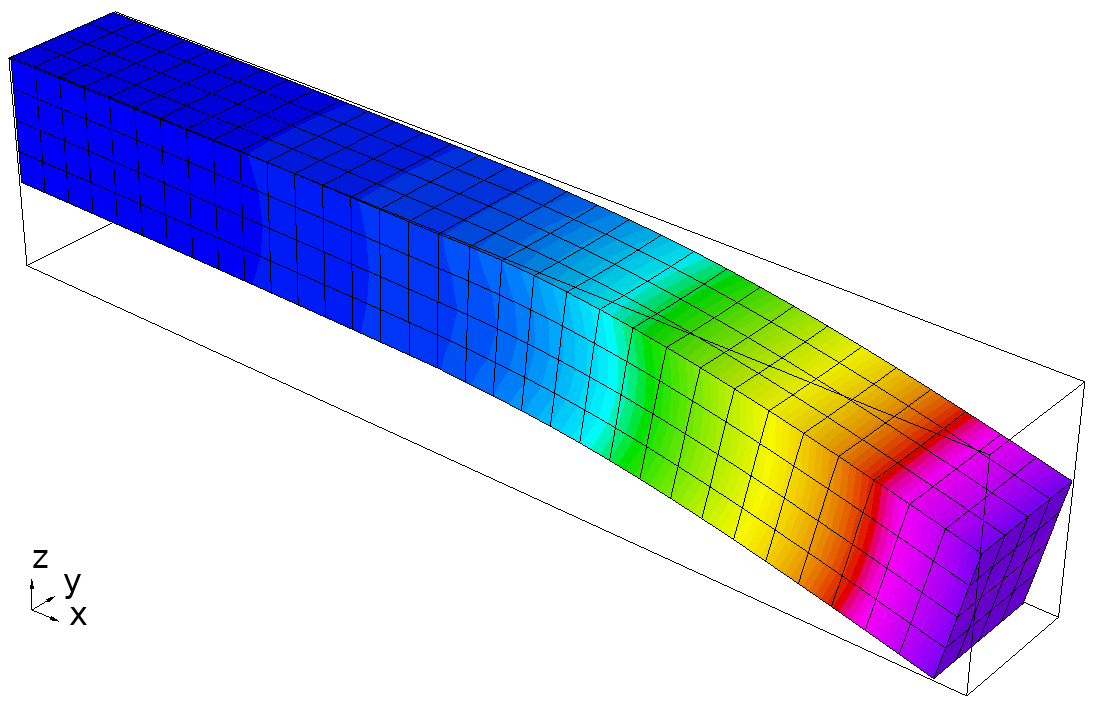}
  \caption{Linear elasticity problem on a 3D beam.}
  \label{fig:mesh_problem_2}
\end{figure}

In the section we consider the solution of the following linear elasticity equation: 
\begin{eqnarray}
\label{eq:test_problem_2}
\mu\Delta u + (\lambda + \mu)\nabla(\nabla\cdot u) &=& f \ \ \text{in} \ \Omega,
\end{eqnarray}
where $\Omega$ is a 3D cantilever beam as shown in Figure~\ref{fig:mesh_problem_2}.
The left end of the beam is fixed, while a constant force (represented by $f$) pulls down 
the beam from the right end. Herein, $u$ is the displacement, while $\lambda$ and $\mu$ 
are the material's Lam\.e constants. 
The initial approximation is again set equal to zero in order to satisfy the 
boundary condition.

Tables \ref{tab:test2_beam_10} and \ref{tab:test2_beam_80} show a comparison between 
\texttt{parGeMSLR} and SchurILU for different uniform mesh refinements obtained using 
first-order Finite Element. For each mesh, the problem becomes more ill-conditioned 
as the ratio $\frac{\lambda}{\mu}$ grows larger. For this reason, we fix $\mu=1$ and 
vary $\lambda=10$ and $\lambda=80$. 
Note that standard AMG converge slowly for this problem since it is almost 
singular. Concisely, \texttt{parGeMSLR} leads to considerable wall-clock time savings 
compared to SchurILU, even when the latter is allowed a higher level of fill-in. 

\begin{table}
\centering
  \caption{Comparison between two-level ILU and the GeMSLR for 3D Linear elasticity problem. $\mu=1$ and $\lambda=10$, Poisson ratio is $\frac{5}{11}\approx 0.455$.}
  \label{tab:test2_beam_10}
  \begin{tabular}{c|cc|ccccc}
    \hline
     prec & size & $n_p$ & k & fill & p-t & i-t & its \\
  \hline
   \multirow{4}{*}{\shortstack{Schur-\\ILU}} 
    & 2,475 & 4 & 
    - & 2.62 & 0.03 & 0.06 & 49 \\
    & 15,795 & 8 & 
    - & 3.78 & 0.32 & 0.60 & 238 \\ 
    & 111,843 & 16 & 
    - & 7.81 & 4.80 & 19.05 & 751 \\
    & 839,619 & 64 & 
    - & 11.82 & 19.67 & - & F \\
   \hline
   \multirow{4}{*}{\shortstack{par\\GeMSLR}} 
    & 2,475 & 4 & 
    20 & 1.94 & 0.12 & 0.01 & 18 \\
    & 15,795 & 8 & 
    40 & 3.58 & 0.92 & 0.04 & 23 \\
    & 111,843 & 16 & 
    40 & 7.86 & 10.06 & 0.64 & 41 \\
    & 839,619 & 64 & 
    80 & 10.05 & 63.25 & 3.13 & 65 \\
  \hline
  \end{tabular}
  % 2475 15795 111843 839619
\end{table}

\begin{table}
\centering
  \caption{Comparison between two-level ILU and the GeMSLR for 3D Linear elasticity problem. $\mu=1$ and $\lambda=80$, Poisson ratio is $\frac{40}{81}\approx 0.494$.}
  \label{tab:test2_beam_80}
  \begin{tabular}{c|cc|ccccc}
    \hline
     prec & size & $n_p$ & k & fill & p-t & i-t & its \\
  \hline
   \multirow{4}{*}{\shortstack{Schur-\\ILU}} 
    & 2,475 & 4 & 
    - & 2.21 & 0.03 & 0.26 & 336 \\
    & 15,795 & 8 & 
    - & 4.03 & 0.35 & 1.48 & 549 \\ 
    & 111,843 & 16 & 
    - & 8.94 & 6.45 & - & F \\
    & 839,619 & 64 & 
    - & 14.75 & 32.17 & - & F \\
   \hline
   \multirow{4}{*}{\shortstack{par\\GeMSLR}} 
    & 2,475 & 4 & 
    20 & 1.91 & 0.15 & 0.01 & 41 \\
    & 15,795 & 8 & 
    40 & 3.58 & 1.09 & 0.15 & 75 \\
    & 111,843 & 16 & 
    80 & 6.48 & 16.16 & 1.49 & 93 \\
    & 839,619 & 64 & 
    120 & 10.31 & 133.2 & 6.15 & 128 \\
  \hline
  \end{tabular}
\end{table}

\subsubsection{Helmholtz equation}

In this section we consider the complex version of \texttt{parGeMSLR} and apply it to 
solve the Helmholtz problem
\begin{equation}
-(\Delta + \omega^2) u = f \ \ \text{in} \ \Omega=[0,1]^3,
\end{equation}
where we use the Perfectly Matched Layer (PML) boundary condition \cite{liu2018solving} and set 
the number of points per wavelength equal to eight.
We used random initial guesses.

\begin{figure}[htbp]
  \centering
  \includegraphics[width=0.95\linewidth]{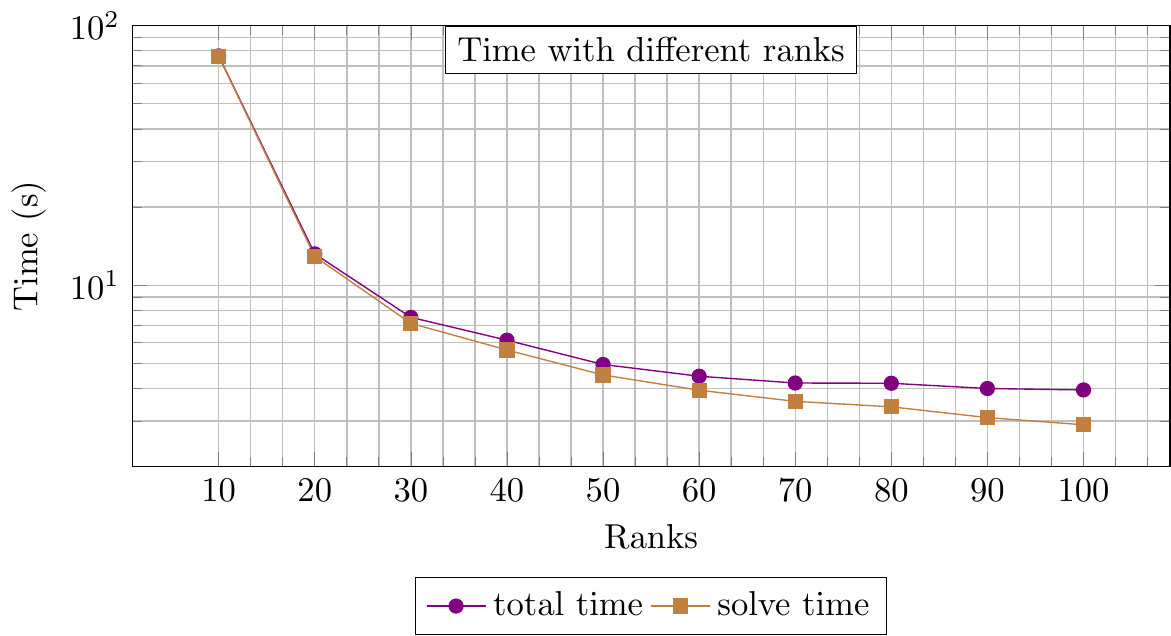}
  \caption{Total and iteration wall-clock times of the 3-level parallel GeMSLR 
  to solve the Helmholtz equation of size $n=50^3$ using 16 MPI processes.}
  \label{fig:hel_50}
\end{figure}

Our first set of experiments focuses on the performance of \texttt{parGeMSLR} where 
the number of low-rank terms is varied as $k=\{10,20,\ldots,100\}$, and the 
number of levels is set equal to $l_{ev}=3$. The size of the Helmholtz 
problem is set equal to $n=50^3$. The maximum fill-in attributed to the low-rank 
correction term was roughly equal to three. Figure~\ref{fig:hel_50} plots 
the parallel wall-clock time as a function of the number of low-rank terms 
$k$ while the number of MPI processes is fixed equal to sixteen. Overall, 
larger values of $k$ lead to lower total and iteration times up to the point 
where the time increase associated with constructing the \texttt{parGeMSLR} 
preconditioner outweighs the gains from improving the convergence rate during 
the iterative solution by FGMRES.  
Next, we consider the same problem but this time we add a complex shift equal to 
$0.05i*\sum_i|A_{ii}|/n_A$ during the the ILU factorization of the on-diagonal 
blocks. The same idea was already considered in \cite{erlangga2004class,dillon_hierarchical_2018,osei-kuffuor_preconditioning_2010} 
but this time we 
apply it in the context of distributed-memory computing and make it available in 
\texttt{parGeMSLR}. 
Similarly to the previous references, adding a shift helps creating a more stable ILU for indefinite problems, i.e., see Table~\ref{tab:hel}. 
\begin{table}
\centering
  \caption{\texttt{parGeMSLR} with/without complex shifts for. 
  %The largest problem size is equal to $n=160^3$.
  The problem size is equal to $n=(4\omega/\pi)^3$.
  }
  \label{tab:hel}
  \begin{tabular}{c|ccccccc|ccc}
    \hline
       & \multicolumn{7}{|c|}{with shift} & \multicolumn{3}{c}{without shift} \\
    \hline
     $\omega$ & $n_p$ & k & fill & r-t & p-t & i-t & its & fill & time & its \\
    \hline
    $5\pi$ & 1 & 0 & 3.40 & 0.04 & 0.02 & 0.05 & 9 & 3.80 & 0.12 & 9 \\
    $7.5\pi$ & 1 & 0 & 3.81 & 0.17 & 0.10 & 0.40 & 20 & 4.76 & 6.47 & 241 \\
    $10\pi$ & 2 & 5 & 3.52 & 0.43 & 0.41 & 1.03 & 36 & 4.11 & 15.48 & 449 \\
    $12.5\pi$ & 4 & 5 & 3.79 & 0.70 & 0.58 & 1.50 & 42 & 4.79 & - & F \\
    $15\pi$ & 8 & 10 & 4.16 & 1.25 & 1.20 & 2.33 & 55 & 4.63 & - & F \\
    $20\pi$ & 16 & 10 & 4.40 & 1.51 & 1.29 & 3.51 & 57 & 4.77 & - & F \\
    $40\pi$ & 64 & 20 & 5.49 & 4.87 & 7.84 & 14.43 & 92 & 5.73 & - & F \\
  \hline
\end{tabular}
\end{table}

\subsection{GPU acceleration of the solution phase}\label{sec:pgpu}

The \texttt{parGeMSLR} library can also take advantage of specialized hardware 
such as GPUs to speed-up numerical kernels. The current release of \texttt{parGeMSLR} 
does not support GPU computing during the setup phase of the GeMSLR preconditioner, 
but allows the 
use of GPUs during the application of the GeMSLR preconditioner, i.e., triangular 
substitutions and dense, rectangular matrix-vector multiplications. Nonetheless, 
accelerating the solution phase might still lead to significant reductions in 
the overall wall-clock time, e.g., when we need to solve for multiple right-hand sides.

\begin{figure}[htbp]
  \centering
  \includegraphics[width=0.95\linewidth]{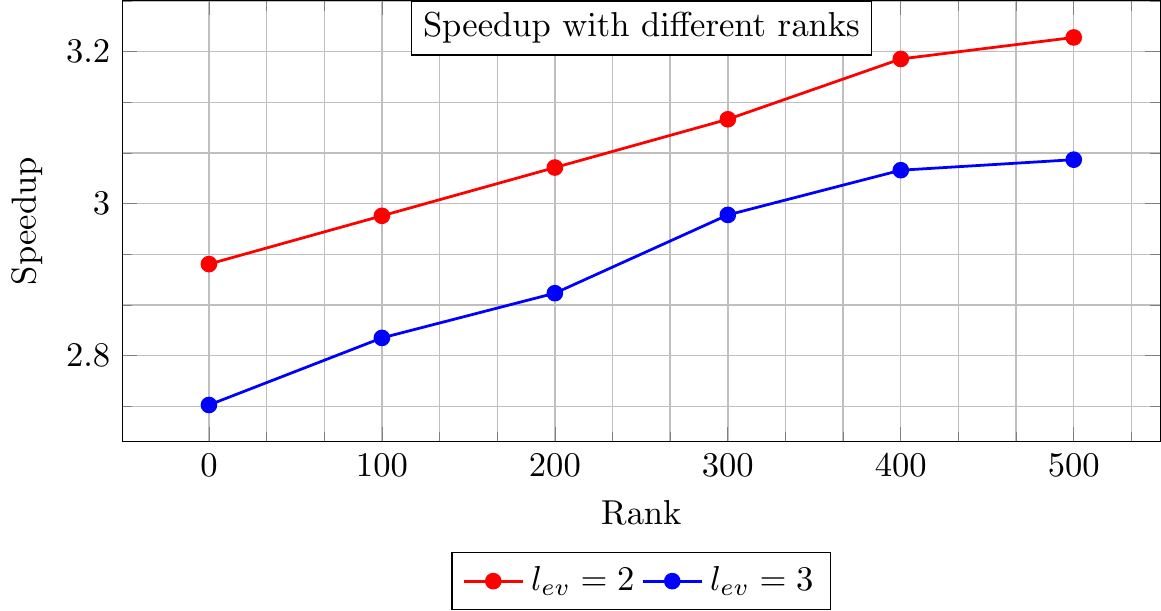}
  \caption{
  Speedup of the solution phase} of \texttt{parGeMSLR} if GPU 
  acceleration is enabled when $l_{ev}=\{2,3\}$, and $k\in \{0,100,200,300,400,500\}$. 
  The problem size is equal to $n=128^3$.
  \label{fig:gpu}
\end{figure}

To demonstrate these benefits, we consider a $n=128^3$ discretization of the model problem 
(\ref{eq:test_problem_1}) and focus on 
the speedup achieved during the solution phase if GPUs are enabled. We set the 
number of levels equal to $l_{ev}=2$ and $l_{ev}=3$, and vary the 
low-rank correction terms 
as $k\in \{0,100,200,300,400,500\}$. At each level, we apply a 4-way partition  
and assign each partition to a separate MPI process binded to a V100 NVIDIA GPU. 
Figure \ref{fig:gpu} plots the speedups achieved by the hybrid CPU+GPU version 
of \texttt{parGeMSLR} during its solve phase. As expected, the peak speedup is obtained 
for the case $k=500$, since the cost to apply the low-rank correction term 
increases linearly with the value of $k$. 

\section{Concluding remarks and future work}\label{sec:conclusion}

In this paper we presented \texttt{parGeMSLR}, a C++ parallel software library 
for the iterative solution of general sparse systems distributed among several 
processor groups communicating via MPI.
environments \cite{dillon_hierarchical_2018}. \texttt{parGeMSLR} is based 
on the GeMSLR preconditioner and can be applied to both real and complex 
systems of linear algebraic equations. The performance of \texttt{parGeMSLR} 
on distributed-memory computing environments was demonstrated on both model 
and real-world problems, verifying the efficiency of the library as a 
general-purpose solver.

As future work we plan to replace standard Arnoldi by either 
its block variant or randomized subspace iteration. This should improve 
performance by reducing latency during the preconditioner setup phase. 
Moreover, the cost of the setup phase can be amortized over the solution 
of linear systems with multiple right-hand sides, e.g., see 
\cite{simoncini1995iterative,hussamenlarged,kalantzis2013accelerating,kalantzis2018scalable}, 
and we plan to apply \texttt{parGeMSLR} to this type of problems. 
In this context, we also plan to apply \texttt{parGeMSLR} to the solution 
of sparse linear systems appearing in eigenvalue solvers 
based on rational filtering \cite{doi:10.1137/20M1349217,xi_computing_2016},  
and domain decomposition \cite{doi:10.1137/17M1154527,kalantzis2020domain}. 

%\bibliographystyle{siamplain}
\bibliographystyle{elsarticle-num}
\bibliography{papers}

\begin{thebibliography}{10}
\expandafter\ifx\csname url\endcsname\relax
  \def\url#1{\texttt{#1}}\fi
\expandafter\ifx\csname urlprefix\endcsname\relax\def\urlprefix{URL }\fi
\expandafter\ifx\csname href\endcsname\relax
  \def\href#1#2{#2} \def\path#1{#1}\fi

\bibitem{saad_iterative_2003}
Y.~Saad, \href{https://doi.org/10.1137/1.9780898718003}{Iterative {Methods} for
  {Sparse} {Linear} {Systems}}, Other {Titles} in {Applied} {Mathematics},
  Society for Industrial and Applied Mathematics, 2003.
\newblock \href {https://doi.org/10.1137/1.9780898718003}
  {\path{doi:10.1137/1.9780898718003}}.

\bibitem{van2003iterative}
H.~A. Van~der Vorst, Iterative Krylov methods for large linear systems, no.~13,
  Cambridge University Press, 2003.

\bibitem{saad_gmres_1986}
Y.~Saad, M.~H. Schultz, \href{https://doi.org/10.1137/0907058}{{GMRES}: {A}
  {Generalized} {Minimal} {Residual} {Algorithm} for {Solving} {Nonsymmetric}
  {Linear} {Systems}}, SIAM Journal on Scientific and Statistical Computing
  7~(3) (1986) 856--869, publisher: Society for Industrial and Applied
  Mathematics.
\newblock \href {https://doi.org/10.1137/0907058} {\path{doi:10.1137/0907058}}.

\bibitem{ruge_algebraic_1987}
J.~W. Ruge, K.~St\{{\textbackslash}"u\}ben,
  \href{https://doi.org/10.1137/1.9781611971057.ch4}{Algebraic {Multigrid}},
  in: Multigrid {Methods}, Frontiers in {Applied} {Mathematics}, Society for
  Industrial and Applied Mathematics, 1987, pp. 73--130.
\newblock \href {https://doi.org/10.1137/1.9781611971057.ch4}
  {\path{doi:10.1137/1.9781611971057.ch4}}.

\bibitem{henson_boomeramg_2002}
V.~E. Henson, U.~M. Yang,
  \href{http://www.sciencedirect.com/science/article/pii/S0168927401001155}{{BoomerAMG}:
  {A} parallel algebraic multigrid solver and preconditioner}, Applied
  Numerical Mathematics 41~(1) (2002) 155--177.
\newblock \href {https://doi.org/10.1016/S0168-9274(01)00115-5}
  {\path{doi:10.1016/S0168-9274(01)00115-5}}.

\bibitem{cleary2000robustness}
A.~J. Cleary, R.~D. Falgout, V.~E. Henson, J.~E. Jones, T.~A. Manteuffel, S.~F.
  McCormick, G.~N. Miranda, J.~W. Ruge, Robustness and scalability of algebraic
  multigrid, SIAM Journal on Scientific Computing 21~(5) (2000) 1886--1908.

\bibitem{bell2012exposing}
N.~Bell, S.~Dalton, L.~N. Olson, Exposing fine-grained parallelism in algebraic
  multigrid methods, SIAM Journal on Scientific Computing 34~(4) (2012)
  C123--C152.

\bibitem{saad_ilut:_1994}
Y.~Saad,
  \href{http://onlinelibrary.wiley.com/doi/10.1002/nla.1680010405/abstract}{{ILUT}:
  {A} dual threshold incomplete {LU} factorization}, Numerical Linear Algebra
  with Applications 1~(4) (1994) 387--402.
\newblock \href {https://doi.org/10.1002/nla.1680010405}
  {\path{doi:10.1002/nla.1680010405}}.

\bibitem{chow1997experimental}
E.~Chow, Y.~Saad, Experimental study of ilu preconditioners for indefinite
  matrices, Journal of computational and applied mathematics 86~(2) (1997)
  387--414.

\bibitem{ernst_why_2012}
O.~G. Ernst, M.~J. Gander,
  \href{http://link.springer.com/10.1007/978-3-642-22061-6_10}{Why it is
  {Difficult} to {Solve} {Helmholtz} {Problems} with {Classical} {Iterative}
  {Methods}}, in: I.~G. Graham, T.~Y. Hou, O.~Lakkis, R.~Scheichl (Eds.),
  Numerical {Analysis} of {Multiscale} {Problems}, Vol.~83, Springer Berlin
  Heidelberg, Berlin, Heidelberg, 2012, pp. 325--363.
\newblock \href {https://doi.org/10.1007/978-3-642-22061-6_10}
  {\path{doi:10.1007/978-3-642-22061-6_10}}.

\bibitem{helmhotz}
X.~Liu, Y.~Xi, Y.~Saad, M.~V. de~Hoop, Solving the three-dimensional
  high-frequency helmholtz equation using contour integration and polynomial
  preconditioning, SIAM Journal on Matrix Analysis and Applications 41~(1)
  (2020) 58--82.

\bibitem{magolu_preconditioning_2000}
M.~Magolu~monga Made, R.~Beauwens, G.~Warz{\'e}e,
  \href{https://onlinelibrary.wiley.com/doi/abs/10.1002/1099-0887%28200011%2916%3A11%3C801%3A%3AAID-CNM377%3E3.0.CO%3B2-M}{Preconditioning
  of discrete {Helmholtz} operators perturbed by a diagonal complex matrix},
  Communications in Numerical Methods in Engineering 16~(11) (2000) 801--817.
\newblock \href
  {https://doi.org/https://doi.org/10.1002/1099-0887(200011)16:11<801::AID-CNM377>3.0.CO;2-M}
  {\path{doi:https://doi.org/10.1002/1099-0887(200011)16:11<801::AID-CNM377>3.0.CO;2-M}}.

\bibitem{erlangga_comparison_2006}
Y.~A. Erlangga, C.~Vuik, C.~W. Oosterlee,
  \href{http://www.sciencedirect.com/science/article/pii/S016892740500108X}{Comparison
  of multigrid and incomplete {LU} shifted-{Laplace} preconditioners for the
  inhomogeneous {Helmholtz} equation}, Applied Numerical Mathematics 56~(5)
  (2006) 648--666.
\newblock \href {https://doi.org/10.1016/j.apnum.2005.04.039}
  {\path{doi:10.1016/j.apnum.2005.04.039}}.

\bibitem{osei-kuffuor_preconditioning_2010}
D.~Osei-Kuffuor, Y.~Saad,
  \href{http://www.sciencedirect.com/science/article/pii/S0168927409001603}{Preconditioning
  {Helmholtz} linear systems}, Applied Numerical Mathematics 60~(4) (2010)
  420--431.
\newblock \href {https://doi.org/10.1016/j.apnum.2009.09.003}
  {\path{doi:10.1016/j.apnum.2009.09.003}}.

\bibitem{anzt2018parilut}
H.~Anzt, E.~Chow, J.~Dongarra, Parilut---a new parallel threshold ilu
  factorization, SIAM Journal on Scientific Computing 40~(4) (2018) C503--C519.

\bibitem{anzt2019parilut}
H.~Anzt, T.~Ribizel, G.~Flegar, E.~Chow, J.~Dongarra, Parilut-a parallel
  threshold ilu for gpus, in: 2019 IEEE International Parallel and Distributed
  Processing Symposium (IPDPS), IEEE, 2019, pp. 231--241.

\bibitem{chow_fine-grained_2015}
E.~Chow, A.~Patel, \href{https://doi.org/10.1137/140968896}{Fine-{Grained}
  {Parallel} {Incomplete} {LU} {Factorization}}, SIAM Journal on Scientific
  Computing 37~(2) (2015) C169--C193.
\newblock \href {https://doi.org/10.1137/140968896}
  {\path{doi:10.1137/140968896}}.

\bibitem{cai_restricted_1999}
X.-C. Cai, M.~Sarkis, \href{https://doi.org/10.1137/S106482759732678X}{A
  {Restricted} {Additive} {Schwarz} {Preconditioner} for {General} {Sparse}
  {Linear} {Systems}}, SIAM Journal on Scientific Computing 21~(2) (1999)
  792--797.
\newblock \href {https://doi.org/10.1137/S106482759732678X}
  {\path{doi:10.1137/S106482759732678X}}.

\bibitem{hysom_efficient_1999}
D.~Hysom, A.~Pothen, \href{https://doi.org/10.1145/331532.331561}{Efficient
  parallel computation of {ILU}(k) preconditioners}, in: Proceedings of the
  1999 {ACM}/{IEEE} conference on {Supercomputing}, {SC} '99, Association for
  Computing Machinery, New York, NY, USA, 1999, pp. 29--es.
\newblock \href {https://doi.org/10.1145/331532.331561}
  {\path{doi:10.1145/331532.331561}}.

\bibitem{karypis_parallel_1997}
G.~Karypis, V.~Kumar, Parallel {Threshold}-based {ILU} {Factorization}, in:
  Supercomputing, {ACM}/{IEEE} 1997 {Conference}, 1997, pp. 28--28.
\newblock \href {https://doi.org/10.1145/509593.509621}
  {\path{doi:10.1145/509593.509621}}.

\bibitem{saad_bilutm_1999}
Y.~Saad, J.~Zhang, \href{https://doi.org/10.1137/S0895479898341268}{{BILUTM}:
  {A} {Domain}-{Based} {Multilevel} {Block} {ILUT} {Preconditioner} for
  {General} {Sparse} {Matrices}}, SIAM Journal on Matrix Analysis and
  Applications 21~(1) (1999) 279--299, publisher: Society for Industrial and
  Applied Mathematics.
\newblock \href {https://doi.org/10.1137/S0895479898341268}
  {\path{doi:10.1137/S0895479898341268}}.

\bibitem{li_parms:_2003}
Z.~Li, Y.~Saad, M.~Sosonkina,
  \href{http://onlinelibrary.wiley.com/doi/10.1002/nla.325/abstract}{{pARMS}: a
  parallel version of the algebraic recursive multilevel solver}, Numerical
  Linear Algebra with Applications 10~(5-6) (2003) 485--509.
\newblock \href {https://doi.org/10.1002/nla.325} {\path{doi:10.1002/nla.325}}.

\bibitem{nievinski_parallel_2018}
I.~C.~L. NIEVINSKI, M.~SOUZA, P.~GOLDFELD, D.~A. AUGUSTO, J.~R.~P. RODRIGUES,
  L.~M. CARVALHO, Parallel {Implementation} of a {Two}-level {Algebraic}
  {ILU}(k)-based {Domain} {Decomposition} {Preconditioner}, TEMA (SÃ£o
  Carlos) 19 (2018) 59--77, publisher: scielo.
\newblock \href {https://doi.org/10.5540/tema.2018.019.01.0059}
  {\path{doi:10.5540/tema.2018.019.01.0059}}.

\bibitem{dillon_hierarchical_2018}
G.~Dillon, V.~Kalantzis, Y.~Xi, Y.~Saad,
  \href{https://doi.org/10.1137/17M1143320}{A {Hierarchical} {Low} {Rank}
  {Schur} {Complement} {Preconditioner} for {Indefinite} {Linear} {Systems}},
  SIAM Journal on Scientific Computing 40~(4) (2018) A2234--A2252.
\newblock \href {https://doi.org/10.1137/17M1143320}
  {\path{doi:10.1137/17M1143320}}.

\bibitem{mandel2003convergence}
J.~Mandel, C.~R. Dohrmann, Convergence of a balancing domain decomposition by
  constraints and energy minimization, Numerical linear algebra with
  applications 10~(7) (2003) 639--659.

\bibitem{farhat2001feti}
C.~Farhat, M.~Lesoinne, P.~LeTallec, K.~Pierson, D.~Rixen, Feti-dp: a
  dual--primal unified feti method—part i: A faster alternative to the
  two-level feti method, International journal for numerical methods in
  engineering 50~(7) (2001) 1523--1544.

\bibitem{heinlein2021combining}
A.~Heinlein, A.~Klawonn, M.~Lanser, J.~Weber, Combining machine learning and
  adaptive coarse spaces---a hybrid approach for robust feti-dp methods in
  three dimensions, SIAM Journal on Scientific Computing 43~(5) (2021)
  S816--S838.

\bibitem{spillane2014abstract}
N.~Spillane, V.~Dolean, P.~Hauret, F.~Nataf, C.~Pechstein, R.~Scheichl,
  Abstract robust coarse spaces for systems of pdes via generalized
  eigenproblems in the overlaps, Numerische Mathematik 126~(4) (2014) 741--770.

\bibitem{amestoy2000mumps}
P.~R. Amestoy, I.~S. Duff, J.-Y. L’Excellent, J.~Koster, Mumps: a general
  purpose distributed memory sparse solver, in: International Workshop on
  Applied Parallel Computing, Springer, 2000, pp. 121--130.

\bibitem{henon2002pastix}
P.~H{\'e}non, P.~Ramet, J.~Roman, Pastix: a high-performance parallel direct
  solver for sparse symmetric positive definite systems, Parallel Computing
  28~(2) (2002) 301--321.

\bibitem{li_low-rank_2017}
R.~Li, Y.~Saad,
  \href{https://epubs.siam.org/doi/abs/10.1137/16M110486X}{Low-{Rank}
  {Correction} {Methods} for {Algebraic} {Domain} {Decomposition}
  {Preconditioners}}, SIAM Journal on Matrix Analysis and Applications 38~(3)
  (2017) 807--828.
\newblock \href {https://doi.org/10.1137/16M110486X}
  {\path{doi:10.1137/16M110486X}}.

\bibitem{boman2020preconditioner}
E.~G. Boman, L.~Cambier, C.~Chen, E.~Darve, S.~Rajamanickam, R.~S. Tuminaro, A
  preconditioner based on sparsified nested dissection and low-rank
  approximation, in: XXI Householder Symposium on Numerical Linear Algebra,
  2020, p. 128.

\bibitem{benzi_sparse_1998}
M.~Benzi, M.~Tuma, \href{https://doi.org/10.1137/S1064827595294691}{A {Sparse}
  {Approximate} {Inverse} {Preconditioner} for {Nonsymmetric} {Linear}
  {Systems}}, SIAM Journal on Scientific Computing 19~(3) (1998) 968--994.
\newblock \href {https://doi.org/10.1137/S1064827595294691}
  {\path{doi:10.1137/S1064827595294691}}.

\bibitem{chow_approximate_1998}
E.~Chow, Y.~Saad, \href{https://doi.org/10.1137/S1064827594270415}{Approximate
  {Inverse} {Preconditioners} via {Sparse}-{Sparse} {Iterations}}, SIAM Journal
  on Scientific Computing 19~(3) (1998) 995--1023.
\newblock \href {https://doi.org/10.1137/S1064827594270415}
  {\path{doi:10.1137/S1064827594270415}}.

\bibitem{janna_block_2010}
C.~Janna, M.~Ferronato, G.~Gambolati,
  \href{https://doi.org/10.1137/090779760}{A {Block} {FSAI}-{ILU} {Parallel}
  {Preconditioner} for {Symmetric} {Positive} {Definite} {Linear} {Systems}},
  SIAM Journal on Scientific Computing 32~(5) (2010) 2468--2484, publisher:
  Society for Industrial and Applied Mathematics.
\newblock \href {https://doi.org/10.1137/090779760}
  {\path{doi:10.1137/090779760}}.

\bibitem{anzt2018incomplete}
H.~Anzt, T.~K. Huckle, J.~Br{\"a}ckle, J.~Dongarra, Incomplete sparse
  approximate inverses for parallel preconditioning, Parallel Computing 71
  (2018) 1--22.

\bibitem{grote1997parallel}
M.~J. Grote, T.~Huckle, Parallel preconditioning with sparse approximate
  inverses, SIAM Journal on Scientific Computing 18~(3) (1997) 838--853.

\bibitem{ye_preconditioning_2019}
X.~Ye, Y.~Xi, Y.~Saad, Preconditioning via gmres in polynomial space (2019).

\bibitem{cai_smash_2018}
D.~Cai, E.~Chow, L.~Erlandson, Y.~Saad, Y.~Xi,
  \href{https://onlinelibrary.wiley.com/doi/abs/10.1002/nla.2204}{{SMASH}:
  {Structured} matrix approximation by separation and hierarchy}, Numerical
  Linear Algebra with Applications 25~(6) (2018) e2204.
\newblock \href {https://doi.org/https://doi.org/10.1002/nla.2204}
  {\path{doi:https://doi.org/10.1002/nla.2204}}.

\bibitem{hackbusch_sparse_1999}
W.~Hackbusch, \href{https://doi.org/10.1007/s006070050015}{A {Sparse} {Matrix}
  {Arithmetic} {Based} on \${\textbackslash}{Cal} {H}\$-{Matrices}. {Part} {I}:
  {Introduction} to \$\{{\textbackslash}{Cal} {H}\}\$-{Matrices}}, Computing
  62~(2) (1999) 89--108.
\newblock \href {https://doi.org/10.1007/s006070050015}
  {\path{doi:10.1007/s006070050015}}.

\bibitem{hackbusch_sparse_2000}
W.~Hackbusch, B.~N. Khoromskij, \href{https://doi.org/10.1007/PL00021408}{A
  {Sparse} \${\textbackslash}{Cal} {H}\$-{Matrix} {Arithmetic}. {Part} {II}:
  {Application} to {Multi}-{Dimensional} {Problems}}, Computing 64~(1) (2000)
  21--47.
\newblock \href {https://doi.org/10.1007/PL00021408}
  {\path{doi:10.1007/PL00021408}}.

\bibitem{xi_superfast_2014}
Y.~Xi, J.~Xia, S.~Cauley, V.~Balakrishnan,
  \href{https://doi.org/10.1137/120895755}{Superfast and {Stable} {Structured}
  {Solvers} for {Toeplitz} {Least} {Squares} via {Randomized} {Sampling}}, SIAM
  Journal on Matrix Analysis and Applications 35~(1) (2014) 44--72, publisher:
  Society for Industrial and Applied Mathematics.
\newblock \href {https://doi.org/10.1137/120895755}
  {\path{doi:10.1137/120895755}}.

\bibitem{chen2018distributed}
C.~Chen, H.~Pouransari, S.~Rajamanickam, E.~G. Boman, E.~Darve, A
  distributed-memory hierarchical solver for general sparse linear systems,
  Parallel Computing 74 (2018) 49--64.

\bibitem{li_schur_2016}
R.~Li, Y.~Xi, Y.~Saad,
  \href{https://onlinelibrary.wiley.com/doi/abs/10.1002/nla.2051}{Schur
  complement-based domain decomposition preconditioners with low-rank
  corrections}, Numerical Linear Algebra with Applications 23~(4) (2016)
  706--729.
\newblock \href {https://doi.org/10.1002/nla.2051}
  {\path{doi:10.1002/nla.2051}}.

\bibitem{falgout_hypre_2002}
R.~D. Falgout, U.~M. Yang, hypre: {A} {Library} of {High} {Performance}
  {Preconditioners}, in: P.~M.~A. Sloot, A.~G. Hoekstra, C.~J.~K. Tan, J.~J.
  Dongarra (Eds.), Computational {Science} {ICCS} 2002, Lecture {Notes} in
  {Computer} {Science}, Springer, Berlin, Heidelberg, 2002, pp. 632--641.
\newblock \href {https://doi.org/10.1007/3-540-47789-6_66}
  {\path{doi:10.1007/3-540-47789-6_66}}.

\bibitem{paralution}
P.~Labs, Paralution v1.1.0, \url{http://www.paralution.com/} (2016).

\bibitem{rupp2016viennacl}
K.~Rupp, P.~Tillet, F.~Rudolf, J.~Weinbub, A.~Morhammer, T.~Grasser, A.~Jungel,
  S.~Selberherr, Viennacl---linear algebra library for multi-and many-core
  architectures, SIAM Journal on Scientific Computing 38~(5) (2016) S412--S439.

\bibitem{hiflow3}
S.~Gawlok, P.~Gerstner, S.~Haupt, V.~Heuveline, J.~Kratzke, P.~Lösel, K.~Mang,
  M.~Schmidtobreick, N.~Schoch, N.~Schween, J.~Schwegler, C.~Song, M.~Wlotzka,
  \href{https://journals.ub.uni-heidelberg.de/index.php/emcl-pp/article/view/42879}{Hiflow3
  – technical report on release 2.0}, Preprint Series of the Engineering
  Mathematics and Computing Lab (EMCL) 0~(06) (2017).
\newblock \href {https://doi.org/10.11588/emclpp.2017.06.42879}
  {\path{doi:10.11588/emclpp.2017.06.42879}}.

\bibitem{balay2001petsc}
S.~Balay, K.~Buschelman, W.~D. Gropp, D.~Kaushik, M.~G. Knepley, L.~C. McInnes,
  B.~F. Smith, H.~Zhang, Petsc, See http://www. mcs. anl. gov/petsc (2001).

\bibitem{trilinos-website}
T.~{T}rilinos~{P}roject {T}eam, The {T}rilinos {P}roject {W}ebsite.

\bibitem{ghysels2017robust}
P.~Ghysels, S.~L. Xiaoye, C.~Gorman, F.-H. Rouet, A robust parallel
  preconditioner for indefinite systems using hierarchical matrices and
  randomized sampling, in: 2017 IEEE International Parallel and Distributed
  Processing Symposium (IPDPS), IEEE, 2017, pp. 897--906.

\bibitem{ghysels2016efficient}
P.~Ghysels, X.~S. Li, F.-H. Rouet, S.~Williams, A.~Napov, An efficient
  multicore implementation of a novel hss-structured multifrontal solver using
  randomized sampling, SIAM Journal on Scientific Computing 38~(5) (2016)
  S358--S384.

\bibitem{rouet2016distributed}
F.-H. Rouet, X.~S. Li, P.~Ghysels, A.~Napov, A distributed-memory package for
  dense hierarchically semi-separable matrix computations using randomization,
  ACM Transactions on Mathematical Software (TOMS) 42~(4) (2016) 1--35.

\bibitem{lidemmel03}
X.~S. Li, J.~W. Demmel, {SuperLU\_DIST}: A scalable distributed-memory sparse
  direct solver for unsymmetric linear systems, ACM Trans. Mathematical
  Software 29~(2) (2003) 110--140.

\bibitem{grigori:hal-01017448}
L.~Grigori, F.~Nataf, S.~Yousef,
  \href{https://hal.inria.fr/hal-01017448}{{Robust algebraic Schur complement
  preconditioners based on low rank corrections}}, Research Report RR-8557,
  {INRIA} (Jul. 2014).

\bibitem{xi_algebraic_2016}
Y.~Xi, R.~Li, Y.~Saad,
  \href{https://epubs.siam.org/doi/abs/10.1137/15M1021830}{An {Algebraic}
  {Multilevel} {Preconditioner} with {Low}-{Rank} {Corrections} for {Sparse}
  {Symmetric} {Matrices}}, SIAM Journal on Matrix Analysis and Applications
  37~(1) (2016) 235--259.
\newblock \href {https://doi.org/10.1137/15M1021830}
  {\path{doi:10.1137/15M1021830}}.

\bibitem{rajamanickam2012shylu}
S.~Rajamanickam, E.~G. Boman, M.~A. Heroux, Shylu: A hybrid-hybrid solver for
  multicore platforms, in: 2012 IEEE 26th International Parallel and
  Distributed Processing Symposium, IEEE, 2012, pp. 631--643.

\bibitem{daas2021two}
H.~A. Daas, T.~Rees, J.~Scott, Two-level nystr$\backslash$" om--schur
  preconditioner for sparse symmetric positive definite matrices, arXiv
  preprint arXiv:2101.12164 (2021).

\bibitem{karypis_fast_1998}
G.~Karypis, V.~Kumar, \href{https://doi.org/10.1137/S1064827595287997}{A {Fast}
  and {High} {Quality} {Multilevel} {Scheme} for {Partitioning} {Irregular}
  {Graphs}}, SIAM Journal on Scientific Computing 20~(1) (1998) 359--392,
  publisher: Society for Industrial and Applied Mathematics.
\newblock \href {https://doi.org/10.1137/S1064827595287997}
  {\path{doi:10.1137/S1064827595287997}}.

\bibitem{catalyurek_hypergraph-partitioning-based_1999}
U.~V. Catalyurek, C.~Aykanat, Hypergraph-partitioning-based decomposition for
  parallel sparse-matrix vector multiplication, IEEE Transactions on Parallel
  and Distributed Systems 10~(7) (1999) 673--693.
\newblock \href {https://doi.org/10.1109/71.780863}
  {\path{doi:10.1109/71.780863}}.

\bibitem{CHACO}
B.~Hendrickson, R.~Leland,
  \href{ftp://ftp.cs.sandia.gov/pub/papers/bahendr/guide.ps.gz}{The {Chaco}
  User's Guide Version 2}, Sandia National Laboratories, Albuquerque NM (1994).

\bibitem{SCOTCH}
F.~Pellegrini,
  \href{{http://gforge.inria.fr/docman/view.php/248/7104/scotch\_user5.1.pdf}}{{
  Scotch} and { libScotch} 5.1 {U}ser's Guide}, {INRIA} Bordeaux Sud-Ouest, IPB
  \& LaBRI, UMR CNRS 5800 (2010).

\bibitem{henon2006parallel}
P.~H{\'e}non, Y.~Saad, A parallel multistage ilu factorization based on a
  hierarchical graph decomposition, SIAM Journal on Scientific Computing 28~(6)
  (2006) 2266--2293.

\bibitem{amestoy_approximate_1996}
P.~R. Amestoy, T.~A. Davis, I.~S. Duff,
  \href{https://doi.org/10.1137/S0895479894278952}{An {Approximate} {Minimum}
  {Degree} {Ordering} {Algorithm}}, SIAM Journal on Matrix Analysis and
  Applications 17~(4) (1996) 886--905, publisher: Society for Industrial and
  Applied Mathematics.
\newblock \href {https://doi.org/10.1137/S0895479894278952}
  {\path{doi:10.1137/S0895479894278952}}.

\bibitem{george_computer_1981}
A.~George, J.~W. Liu, Computer {Solution} of {Large} {Sparse} {Positive}
  {Definite}, Prentice Hall Professional Technical Reference, 1981.

\bibitem{bienz2019node}
A.~Bienz, W.~D. Gropp, L.~N. Olson, Node aware sparse matrix--vector
  multiplication, Journal of Parallel and Distributed Computing 130 (2019)
  166--178.

\bibitem{mfem}
R.~Anderson, J.~Andrej, A.~Barker, J.~Bramwell, J.-S. Camier, J.~C.~V. Dobrev,
  Y.~Dudouit, A.~Fisher, T.~Kolev, W.~Pazner, M.~Stowell, V.~Tomov,
  I.~Akkerman, J.~Dahm, D.~Medina, S.~Zampini, {MFEM}: A modular finite element
  library, Computers \& Mathematics with Applications (2020).
\newblock \href {https://doi.org/10.1016/j.camwa.2020.06.009}
  {\path{doi:10.1016/j.camwa.2020.06.009}}.

\bibitem{mfem-web}
{MFEM}: Modular finite element methods {[Software]}, \url{mfem.org}.
\newblock \href {https://doi.org/10.11578/dc.20171025.1248}
  {\path{doi:10.11578/dc.20171025.1248}}.

\bibitem{glvis-tool}
{GLVis}: Opengl finite element visualization tool, \url{glvis.org}.
\newblock \href {https://doi.org/10.11578/dc.20171025.1249}
  {\path{doi:10.11578/dc.20171025.1249}}.

\bibitem{baker_multigrid_2011}
A.~H. Baker, R.~D. Falgout, T.~V. Kolev, U.~M. Yang,
  \href{https://doi.org/10.1137/100798806}{Multigrid {Smoothers} for
  {Ultraparallel} {Computing}}, SIAM Journal on Scientific Computing 33~(5)
  (2011) 2864--2887, publisher: Society for Industrial and Applied Mathematics.
\newblock \href {https://doi.org/10.1137/100798806}
  {\path{doi:10.1137/100798806}}.

\bibitem{liu2018solving}
X.~Liu, Y.~Xi, Y.~Saad, M.~V. de~Hoop, Solving the 3d high-frequency helmholtz
  equation using contour integration and polynomial preconditioning, arXiv
  preprint arXiv:1811.12378 (2018).

\bibitem{erlangga2004class}
Y.~A. Erlangga, C.~Vuik, C.~W. Oosterlee, On a class of preconditioners for
  solving the helmholtz equation, Applied Numerical Mathematics 50~(3-4) (2004)
  409--425.

\bibitem{simoncini1995iterative}
V.~Simoncini, E.~Gallopoulos, An iterative method for nonsymmetric systems with
  multiple right-hand sides, SIAM Journal on Scientific Computing 16~(4) (1995)
  917--933.

\bibitem{hussamenlarged}
A.~Hussam, L.~GRIGORI, P.~H{\'e}non, P.~RICOUX, Enlarged gmres for solving
  linear systems with one or multiple right-hand sides.

\bibitem{kalantzis2013accelerating}
V.~Kalantzis, C.~Bekas, A.~Curioni, E.~Gallopoulos, Accelerating data
  uncertainty quantification by solving linear systems with multiple right-hand
  sides, Numerical Algorithms 62~(4) (2013) 637--653.

\bibitem{kalantzis2018scalable}
V.~Kalantzis, A.~C.~I. Malossi, C.~Bekas, A.~Curioni, E.~Gallopoulos, Y.~Saad,
  A scalable iterative dense linear system solver for multiple right-hand sides
  in data analytics, Parallel Computing 74 (2018) 136--153.

\bibitem{doi:10.1137/20M1349217}
V.~Kalantzis, Y.~Xi, L.~Horesh, \href{https://doi.org/10.1137/20M1349217}{Fast
  randomized non-hermitian eigensolvers based on rational filtering and matrix
  partitioning}, SIAM Journal on Scientific Computing 43~(5) (2021) S791--S815.
\newblock \href {http://arxiv.org/abs/https://doi.org/10.1137/20M1349217}
  {\path{arXiv:https://doi.org/10.1137/20M1349217}}, \href
  {https://doi.org/10.1137/20M1349217} {\path{doi:10.1137/20M1349217}}.

\bibitem{xi_computing_2016}
Y.~Xi, Y.~Saad,
  \href{https://epubs.siam.org/doi/abs/10.1137/16M1061965}{Computing {Partial}
  {Spectra} with {Least}-{Squares} {Rational} {Filters}}, SIAM Journal on
  Scientific Computing 38~(5) (2016) A3020--A3045.
\newblock \href {https://doi.org/10.1137/16M1061965}
  {\path{doi:10.1137/16M1061965}}.

\bibitem{doi:10.1137/17M1154527}
V.~Kalantzis, Y.~Xi, Y.~Saad, \href{https://doi.org/10.1137/17M1154527}{Beyond
  automated multilevel substructuring: Domain decomposition with rational
  filtering}, SIAM Journal on Scientific Computing 40~(4) (2018) C477--C502.
\newblock \href {http://arxiv.org/abs/https://doi.org/10.1137/17M1154527}
  {\path{arXiv:https://doi.org/10.1137/17M1154527}}, \href
  {https://doi.org/10.1137/17M1154527} {\path{doi:10.1137/17M1154527}}.

\bibitem{kalantzis2020domain}
V.~Kalantzis, A domain decomposition rayleigh--ritz algorithm for symmetric
  generalized eigenvalue problems, SIAM Journal on Scientific Computing 42~(6)
  (2020) C410--C435.

\end{thebibliography}

\appendix

\end{document}